\documentclass[lettersize,journal]{IEEEtran}
\usepackage{amsmath,amsfonts}
\usepackage{algorithmic}
\usepackage{algorithm}
\usepackage{array}
\usepackage[caption=false,font=normalsize,labelfont=sf,textfont=sf]{subfig}
\usepackage{textcomp}
\usepackage{stfloats}
\usepackage{url}
\usepackage{verbatim}
\usepackage{graphicx}
\usepackage{cite}
\usepackage[hidelinks]{hyperref}
\usepackage{listings}
\usepackage{xcolor}

\begin{document}

\title{One Prompt, Many Sounds: Modeling Listener Variability in LLM-Based Equalization}

\author{Ioannis Stylianou, Jon Francombe, Pablo Mart\'inez-Nuevo, Sven Ewan Shepstone,~\IEEEmembership{Member,~IEEE}, and Zheng-Hua Tan,~\IEEEmembership{Senior Member,~IEEE}%
    \thanks{Ioannis Stylianou, Jon Francombe, Pablo Martínez-Nuevo and Sven Ewan Shepstone are with Bang \& Olufsen A/S, Struer, Denmark. (e-mail: {iost,jofr,pmn,ssh})@bang-olufsen.dk.}%
    \thanks{Ioannis Stylianou and Zheng-Hua Tan are with the Department of Electronic Systems, Aalborg University, Aalborg, Denmark. (e-mail: {iost,zt})@es.aau.dk. Zheng-Hua Tan is also with the Pioneer Centre for AI, Copenhagen, Denmark.}%
}

\markboth{JOURNAL OF SELECTED TOPICS IN SIGNAL PROCESSING}%
{Shell \MakeLowercase{\textit{et al.}}: A Sample Article Using IEEEtran.cls for IEEE Journals}


\maketitle

\begin{abstract}
Conventional audio equalization is a static process that requires manual and cumbersome adjustments to adapt to changing listening contexts (e.g., mood, location, or social setting). In this paper, we introduce a Large Language Model (LLM)-based alternative that maps natural language text prompts to equalization settings. This enables a conversational approach to sound system control. By utilizing data collected from a controlled listening experiment, our models exploit in-context learning and parameter-efficient fine-tuning techniques to reliably align with population-preferred equalization settings. Our evaluation methods, which leverage distributional metrics that capture users' varied preferences, show statistically significant improvements in distributional alignment over random sampling and static preset baselines. These results indicate that LLMs could function as ``artificial equalizers," contributing to the development of more accessible, context-aware, and expert-level audio tuning methods.

\end{abstract}


\section{Introduction}
\IEEEPARstart{R}{ecommender} systems that suggest what content to consume are well established (e.g., Netflix movies or Spotify tracks recommendations \cite{gomez2015netflix, 10.1145/2959100.2959120}), and they learn from massive datasets. Instead of determining \emph{what} to recommend, in this paper we focus on \emph{how} audio content should be reproduced by choosing from a potentially infinite set of parameters, in a domain where data is extremely scarce. We begin by focusing specifically on the frequency response as a candidate loudspeaker parameter, and on how it should adapt to natural language queries.

Digital equalization typically involves implementing filters between the signal source and the system being equalized to achieve a target frequency response. Well-designed equalizers can significantly improve loudspeaker performance, and manufacturers often rely on digital equalization to achieve the desired frequency response \cite{dziechcinski2005comparison}. 

However, factors such as phase distortion and environmental variability contribute to the non-trivial nature of audio equalization, necessitating sophisticated approaches to achieve high-quality sound reproduction \cite{valimaki2016all}. Recent advances in smart equalization have increasingly incorporated artificial intelligence and machine learning techniques to automate and enhance the equalization process. A significant development in this area is the automatic selection of equalization parameters based on intelligent analysis of the audio signal \cite{pepe2022deep, pepe2020designing} or by end-to-end equalization matching \cite{martinez2018end}. Nevertheless, many of these approaches primarily focus on optimizing the audio signal directly, often without taking into account the listening context or individual preferences.

The listening context remains a persistent challenge in audio equalization, exemplified by the semantic gap between audio engineers and non-technical consumers. A core difficulty lies not only in the lack of technical vocabulary \cite{pardo2012building} but in the inherent subjectivity and ambiguity of natural language. A descriptor such as ``warm" or ``clear" does not correspond to a single, universal frequency response; rather, it represents a distribution of preferences that varies across individuals and cultures \cite{kramsch2014language}. Existing methods often reduce these descriptors to single point-estimates, failing to capture the full nuance and diversity of user intent. In this work, we seek to address these limitations by developing an approach that explicitly models this subjectivity, treating the mapping from language to acoustic parameters as a one-to-many problem.

Traditional techniques for loudspeaker configuration procedures may still not be optimal since these processes are obtrusive and static, i.e., they require manual in-situ measurements, microphone measurements, and/or user input when changes occur in the physical setup \cite{goldberg2003statistical}. Moreover, this approach ignores users’ preferences by essentially only considering expert informed preferences (i.e., of the loudspeaker designers). Context-specific reproduction is also neglected; for example, higher bass and volume levels may be preferred at a house party as opposed to dinner time. Furthermore, since different users may interpret commands differently, a rigid rule-based system is insufficient. We therefore develop an artificial intelligence-based framework that leverages Large Language Models (LLMs) to generate predictions based on the distribution of likely equalization settings given a text prompt, thereby acknowledging and preserving the diversity of human perception.

Our work presents the following key contributions.

\begin{itemize}
    \item \textbf{LLM-Based Recommender:} A novel framework that leverages LLMs to interpret audio descriptions and generate plausible equalization adjustments, rather than deterministic point estimates.
    \item \textbf{Evaluation Framework:} A methodology that employs distributional metrics to rigorously assess how well the model captures the variance and spread of human subjective responses.
    \item \textbf{Dataset:} A curated dataset pairing natural language prompts with appropriate loudspeaker equalization settings, gathered from a controlled listening experiment involving 11 participants.
\end{itemize}

This paper is organized as follows. Section \ref{sec:Related Work} delves deeper into the related work. Section \ref{sec:Experiment} details the data and the listening experiment conducted for data collection. Section \ref{sec:LLMs} provides an outline of the different LLM methods explored and how they can potentially enhance the quality of audio adjustments. Section \ref{sec:Evaluation} presents insights on the evaluation process and the metrics utilized. Section \ref{sec:Results} presents the recommendation quality of the different methods explored, based on the developed evaluation metrics, and finally, Section \ref{sec:Conclusion} presents the conclusion of this work and directions for future research.

\section{Related Work} \label{sec:Related Work}
Our work lies at the intersection of intelligent equalization, natural language understanding, and recommender systems. We review prior work in these domains to position our contributions.

\subsection{Automatic and Intelligent Equalization}
The automation of audio equalization has a rich history, evolving from classic signal processing to modern machine learning. Traditional approaches often rely on algorithmic solutions, such as inverting a measured room impulse response using Finite Impulse Response (FIR) or Infinite Impulse Response (IIR) filters \cite{mourjopoulos1985variation}.

Recent advances have leveraged machine learning to create more sophisticated and data-driven equalization systems. For example, Pepe et al. \cite{pepe2022deep, pepe2020designing} employed deep neural networks (multilayer perceptrons, convolutional neural networks, and convolutional autoencoders) that learn to predict optimal equalization parameters by analyzing the spectral content of audio signals. Similarly, Martinez et al.\ \cite{martinez2018end} proposed an end-to-end deep learning model that learns a direct mapping from a distorted audio signal to an equalized one, effectively treating the equalizer as a learned transformation.

However, these approaches are fundamentally driven by the audio signal itself, aiming for an objectively ``correct" or ``clean" output. Our work diverges from this paradigm by using the user's subjective description, expressed in natural language, as the primary input for generating equalization settings, thereby shifting the focus from signal correction to preference fulfillment.

\subsection{Natural Language for Audio Control}
The challenge of bridging the semantic gap between high-level human descriptions and low-level technical parameters is a long-standing problem in audio engineering. Early pioneering work by Pardo, Seetharaman et al.\ \cite{pardo2012building, seetharaman2016audealize} explored mapping single-word semantic descriptors (e.g., ``warm," ``bright," ``muddy") to specific equalization curves. While foundational, these systems relied on fixed vocabularies and could not interpret complex or compositional user requests.

More recently, the field has expanded into broader language-conditioned audio tasks. Large-scale generative models (e.g. AudioLM \cite{borsos2023audiolm}, MusicLM \cite{agostinelli2023musiclm}) synthesize audio from scratch, while instruction-following editing models (e.g. Audit \cite{NEURIPS2023_e1b619a9}, Prompt-guided editing \cite{xu2024prompt}) modify audio representations directly. Closer to our specific domain of parameter control, recent works such as LLM2Fx \cite{doh2025can} and Text2Fx \cite{chu2025text2fx} have demonstrated the capability of LLMs to predict audio effect parameters from natural language.

However, a critical distinction exists between these approaches and our work. Prior methods typically  formulate the task as a regression problem, optimizing for a single set of parameters (a point estimate). This assumes a deterministic mapping where a prompt like ``make it clearer" corresponds to a single ``correct" setting. In contrast, our work posits that such prompts are inherently subjective and polysemous. Instead of predicting a single point, we focus on modeling the distribution of plausible parameters, leveraging the compositional understanding of LLMs to capture the variance in human interpretation rather than collapsing it to a mean. Furthermore, to the best of our knowledge, this is the first study that includes \emph{training} LLMs to predict audio effect parameters from user data.

\subsection{Context-Aware Audio and Subjective Preference}
Context-awareness has become a key feature in modern audio systems. Relevant implementations include adaptive volume control for speech intelligibility in noisy environments, which adjusts playback levels based on noise analysis \cite{felber2011automatic}. Another example is context-aware music recommendation, where signals such as time of day, user activity and geolocation (detected via smartphones) inform dynamic song selection to fit the listener's situational state \cite{lozano2021context}. This concept of user-profiling extends beyond audio, with researchers utilizing user recordings to inform broader multimedia recommendations \cite{shepstone2013audio}. 

The ultimate goal of such systems is to align audio reproduction with the user's situation. However, most existing context-aware systems define ``context" in purely physical terms. They can adapt to a ``noisy" environment but cannot capture the user's subjective intent within that context (e.g., \emph{``I'm having a party and want more bass"} vs.\ \emph{``I'm listening to a podcast and need vocal clarity"}). This leads to a static or rule-based adaptation that may not align with the user's desired experience.

Our framework redefines context-awareness by treating the user's natural language request as the primary source of contextual information. This allows for a dynamic adaptation that directly models the distribution of user preferences, allowing the system to propose settings that reflect the diversity of human perception rather than relying solely on expert-defined presets.


\section{Data Collection} \label{sec:Experiment}
\begin{figure}
    \begin{center}
    \includegraphics[width=\linewidth]{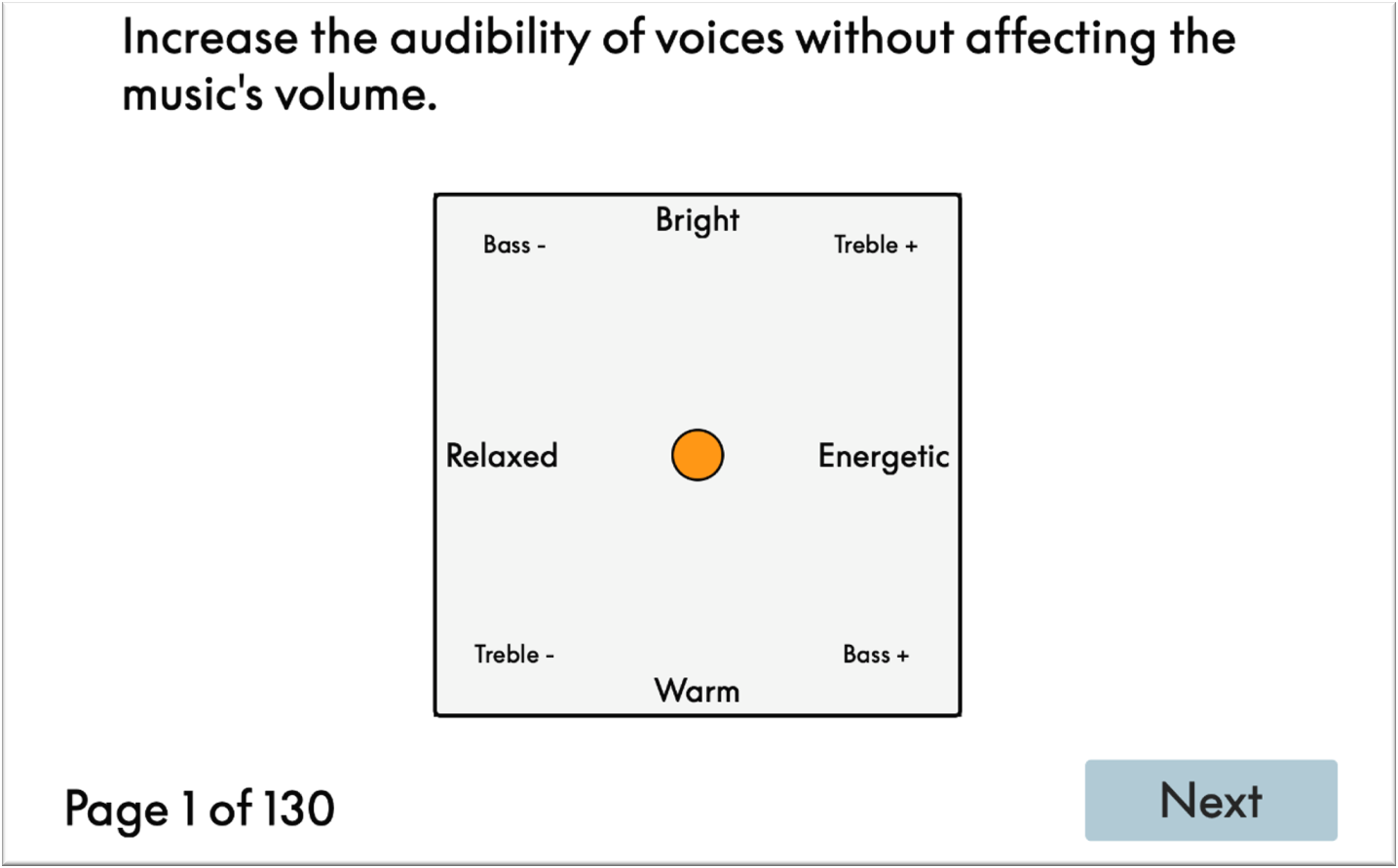}
    \caption{Illustration of the interface used for the data collection experiment. Each participant listened to audio through headphones, interact with the interface through a tablet, and attempt to satisfy the presented sentence through audio adjustments.}
    \label{fig:experiment}
    \end{center}
\end{figure}

We conduct a listening experiment to gather authentic user input regarding how the frequency response should be adjusted in response to a given command. This command is conveyed as text, which we refer to as a \textbf{``prompt."} The experiment deliberately recruits participants who are not necessarily experts in audio engineering, thereby capturing the responses of a general audience. Consequently, we have streamlined their role by introducing a simple 2D equalization controller, that displays guiding text at the borders of the UI square, as shown in Figure \ref{fig:experiment}. More information about this controller is provided in Subsection \ref{subsec:controller}.

\subsection{Process}
The data collection process begins with acquiring a diverse set of prompts, which accurately reflect real user-machine interactions. We gather these prompts through a five-question survey. The first question requires the participant to recall an enjoyable audio experience. Similarly, the second question inquires about an experience that was disliked, the third about an experience where some change was desirable, the fourth about an occasion where something specific could be singled out from the audio, and lastly, an abstract question about how the environment affects sound. The questionnaire was answered by a total of 20 participants and the resulting answers provide descriptions of personalized audio experiences. Rather than automating the prompt extraction process from these responses, we manually review each narrative. Based on these underlying concepts, we formulate several plausible natural-language commands that a user might realistically issue to a conversational recommender system to achieve that specific experience. Figure \ref{fig:json_jstsp} illustrates an example of this.

\begin{figure}[t]
\centering
\begin{lstlisting}[
  basicstyle=\ttfamily\footnotesize,
  breaklines=true,
  columns=fullflexible,
  frame=single
]
[
  ...,
  {
    "Questionnaire Response": "I was listening to an acapella cover of a song and I loved that I could pick out all the interesting harmonies",
    "Extracted Prompts": [
      "Enhance the clarity of harmonies in the audio.",
      "Ensure individual voices are distinct and identifiable.",
      "Optimize vocal balance for acapella performances.",
      "Preserve the nuances of the song's harmonies.",
      "Adjust the sound to emphasize the intricacies of vocal harmonization."
    ]
  },
  ...
]
\end{lstlisting}
\caption{Example JSON snippet from the collected questionnaire data.}
\label{fig:json_jstsp}
\end{figure}

The next stage of the data collection process is audio selection. While the audio samples will not directly serve as input for the models, they are essential for guiding the participants during the labeling process. To ensure comprehensive coverage of various domestic audio scenarios, the pool of audio sources is selected to be as diverse and extensive as possible. Examples of domestic audio scenarios include music, movies, podcasts, radio shows, and audiobooks. Incorporating a wide range of audio content helps to ensure that the dataset represents the variety of sounds typically encountered in domestic environments. The details about the audio types, sources and licensing are presented in Table \ref{tab:audio_sources}.

\begin{table}[!t]
  \caption{Audio Sources, Types, and Licenses}
  \label{tab:audio_sources}
  \centering
  \begin{tabular}{|c|c|l|}
    \hline
    \textbf{Source} & \textbf{Audio Type} & \textbf{Licence} \\
    \hline
    Frontify & Music 1 & Custom \\
    Free Music Archive & Music 2 & \href{https://creativecommons.org/licenses/by-sa/4.0/}{CC BY-SA 4.0} \\
    Blender Studio & Movies 1 & \href{https://creativecommons.org/licenses/by/4.0/}{CC BY 4.0} \\
    Public Domain Movies & Movies 2 & \href{https://creativecommons.org/public-domain/cc0/}{CC0 Public Domain} \\
    Pixabay & Nature sounds / Noise & \href{https://pixabay.com/service/license-summary/}{Pixabay Licence} \\
    LibriSpeech & Audiobooks & \href{https://creativecommons.org/licenses/by/4.0/}{CC BY 4.0} \\
    \hline
  \end{tabular}
\end{table}

With both prompt and audio sets collected, the following step of the process involves combining them into a dataset that will be used for the listening experiment. A straightforward approach would involve generating all possible combinations of the two sets. However, this would lead to a vast number of combinations, many of which would be implausible in real-world scenarios. For instance, combining a prompt like \textit{I can’t hear very clearly what the narrator is saying,} with instrumental music would create an illogical entry. 

Furthermore, since human labeling is required for this task, we are limited in the number of entries we can feasibly label. As a result, the final number of labeled entries is significantly smaller than the number of possible combinations.

Therefore, the feature engineering process is carried out manually while ensuring that the selected combinations are both realistic and relevant to the objectives of the dataset. By carefully considering the context of each audio type and prompt, we aim to create entries that reflect plausible real-world scenarios. Additionally, care is taken to balance the dataset across different audio types and prompt categories to ensure comprehensive coverage of the problem space. The resulting set contains 120 pairs of unique prompts and audio excerpts. To provide insight into the linguistic diversity of the dataset, we present key statistics in Table \ref{tab:prompt_stats}. The prompts exhibit a vocabulary size of 380 unique words and an average length of 9.89 words. This length includes all function words and filler words, as no text preprocessing, stop-word removal, or stemming was applied prior to the length calculation, ensuring that the natural phrasing of the user was maintained. Crucially, the dataset captures varying degrees of semantic ambiguity; some prompts yield highly consistent user settings (low generalized variance), while ambiguous prompts result in high generalized variance (Figure \ref{fig:responses}), justifying the need for the distributional modeling approach described in this work. Section \ref{sec:Evaluation} provides more details on this.

\begin{table}[!t]
  \caption{Statistics of Collected Prompts}
  \label{tab:prompt_stats}
  \centering
  \begin{tabular}{|c|c|}
    \hline
    \textbf{Statistic} & \textbf{Value} \\
    \hline
    Total Prompts & 120 \\
    \hline
    Total Annotations & 1320 \\
    \hline
    Vocabulary Size & 380 \\
    \hline
    Avg. Words per Prompt & 9.89 \\
    \hline
    Avg. Generalized Variance ($|\Sigma|$) & 41.60 \\
    \hline
    Min. Generalized Variance ($|\Sigma|$) & 2.19 \\
    \hline
    Max. Generalized Variance ($|\Sigma|$) & 250.60 \\
    \hline

  \end{tabular}
\end{table}

\subsection{Controller} \label{subsec:controller}
\begin{figure}
    \begin{center}
    \includegraphics[width=\columnwidth]{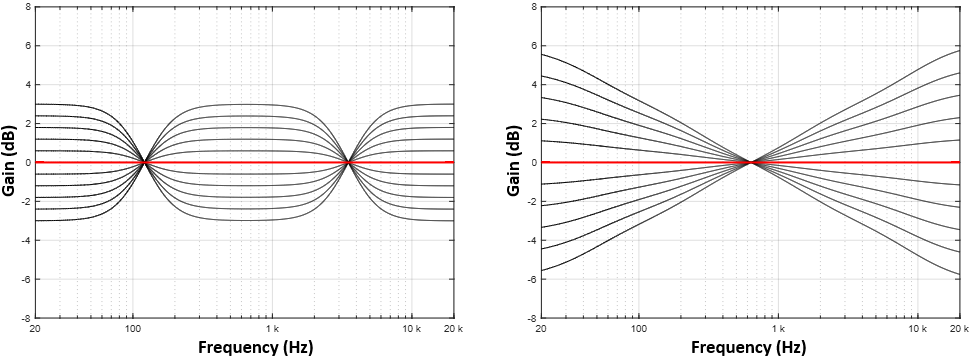}
    \caption{The filters that comprise the equalization controller. A horizontal movement in the interface introduces a ``smile curve" effect (left) while a vertical movement applies a linear adjustment (right).}
    \label{fig:beosonic_filt}
    \end{center}
\end{figure}

After being subjected to the feature engineering process, the features of the dataset are now ready to be labeled. We use a simplified user interface that allows the experiment participants to explore a range of timbral controls with a single, two-dimensional controller, identical to the Beosonic model as implemented in the Beosound Theatre \cite{beosound_theatre_guide}. With this approach, the transfer function that defines the audio system's frequency response is defined by two parameters. The one introduces a ``smile curve" effect, attenuating bass and treble while boosting midrange frequencies at low settings, and the opposite at high ones. The other applies a linear adjustment across the (log-scale) spectrum, boosting bass and attenuating treble at low settings, and vice versa. Figure \ref{fig:beosonic_filt} provides an illustration of the filters. Together, these parameters dynamically shape the audio signal by combining their respective effects to create a customized frequency response. Each parameter is then associated with one dimension of the controller; the ``smile curve" for the x-axis and the linear adjustment for the y-axis, with a gain up of to $\pm$6 dB. The corners of the Beosonic square i.e., [6,6], [-6,6], [-6,-6] and [6,-6], correspond to ``maximum treble", ``minimum bass", ``minimum treble" and ``maximum bass" respectively. This approach offers intuitive control over sound characteristics, balancing ease of use with flexible audio customization.

\subsection{Task}

The participants of the experiment provided their informed consent by signing a consent form. While listening to audio, each one was able to interact with the controller through a tablet. The corresponding prompt was presented for each audio clip, and the users could move the interface's cursor to any position within the square. Guided by the changes on the audio triggered by the cursors movement, the human labeler was requested to select the final position that, in their opinion, best matched the phrase.

During the experiment the 120 prompt-audio entries were presented to each participant, with an extra 10 repeats comprising a case control set. For each prompt-audio pair, responses from 11 participants were collected. This resulted in 120 labels of dimension 11x2 corresponding to 11 participants and the 2 parameters of Beosonic. 

While a participant pool of N = 11 is relatively small for establishing global population norms, this experimental design provides a dense annotation of the 120 prompts. In the context of this study, these 11 responses serve as an empirical distribution derived from the present participant pool for each specific prompt. This allows us to evaluate the model's ability to predict inter-subject disagreement, validating the distributional loss mechanism even if the specific preferences may vary in a larger population.

\section{Methods} \label{sec:LLMs}

With the experiment completed, the collected data can now be utilized to evaluate and refine recommendation quality. The 120 entries are split into 60, 30 and 30 for training, validation and testing respectively. We categorize the approaches that we explore into In-Context Learning (ICL) and Parameter Efficient Fine Tuning (PEFT). We select Phi-3.5-mini-instruct \cite{abdin2024phi3technicalreporthighly} for its \emph{relatively} compact size (3.8B parameters), open-source availability, and rich documentation.

\begin{figure*}
  \centering
  \includegraphics[width=\textwidth]{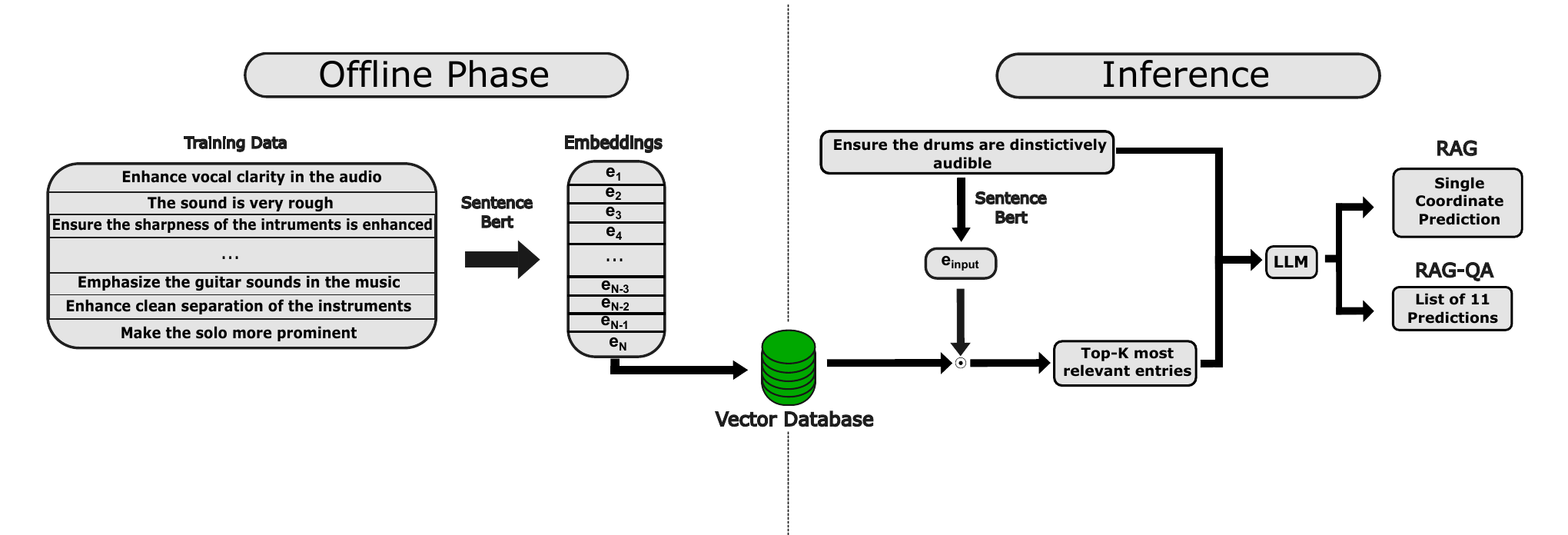}
  \caption{Outline of RAG and RAG-QA recommendation approaches. The sentence similarity is computed as the dot product between the embedding of the input sentence and the embeddings of the database.}
  \label{fig:RAG}
\end{figure*}

\subsection{In-Context Learning}
In this subsection we outline the different approaches for enhancing recommendation quality, without updating any model weights. ICL refers to the ability of large language models to perform new tasks by conditioning on a few input-output examples provided within the input text, without any explicit fine-tuning \cite{brown2020language}. This capability allows the model to smoothly adapt to various tasks during inference by recognizing patterns and making predictions based on the context provided.

\begin{itemize}
    \item \textbf{Text2Beosonic}. A \textbf{zero-shot approach} in which the task is explained in detail, along with an explicit mapping between various audio expressions and the Beosonic space (e.g., ``warm" is mapped to [0, -6]). This mapping, which is provided to the model through the system prompt, is based on automatic text clustering for audio attribute elicitation \cite{francombe2017automatic}. The Text2Beosonic model serves as a baseline to evaluate the effectiveness of providing an LLM with an expert-defined mapping rather than relying on direct training examples. While this approach does not leverage real user feedback, it tests whether a well-explained framework can yield reasonable recommendations without prior exposure to the data.

    \item \textbf{Static-ICL}. A \textbf{few-shot approach}, where a basic explanation of the task is provided, along with a few examples of the training set. This approach helps assess how well the model generalizes with limited labeled examples. By providing a small set of training data, the model can recognize patterns in how natural language is mapped to the Beosonic space. However, this method is limited by the static nature of the examples, which may not always align with diverse user preferences.
\end{itemize}

A more sophisticated approach to ICL is Retrieval Augmented Generation (RAG) \cite{lewis2020retrieval}. RAG enhances the performance of LLMs by integrating external information retrieval into the generative process. It operates through four key stages: indexing external data sources into machine-readable formats, retrieving relevant information based on user queries, augmenting the prompt with this retrieved data, and generating responses that combine the LLM's internal knowledge with the augmented context. The following two recommender systems utilize this method.

\begin{itemize}
    \item \textbf{RAG}. Similar to \textbf{Static-ICL}, this approach presents to the model a task description, but instead of using a fixed set of examples, it dynamically retrieves the most relevant ones. The retrieval process is based on the \textbf{semantic similarity} between the input prompt and prompts from the training set, computed using embeddings from \textbf{Sentence-BERT} \cite{reimers2019sentencebertsentenceembeddingsusing}. \footnote{While newer architectures like Sentence-T5 \cite{ni2022sentence} achieve marginal improvements on broad semantic benchmarks, they typically impose a higher computational overhead. Given our system’s focus on efficient adaptation, Sentence-BERT provides the optimal balance between retrieval accuracy and inference latency. Furthermore, we select a text-only embedding model over multimodal alternatives (e.g., CLAP \cite{elizalde2023clap}) because our retrieval mechanism focuses exclusively on the linguistic nuance of the user's command. Future iterations could incorporate audio content analysis, but for this study, we isolate the text-to-parameter mapping capabilities.}
    
    Given the plethora of audio expressions, a static few-shot approach may not always provide relevant guidance for new prompts. RAG dynamically retrieves the most contextually relevant examples based on semantic similarity, making it more adaptive to variations in user input. This approach ensures that the model’s predictions are informed by examples that are most closely related to the specific audio expression being analyzed.

    \item \textbf{RAG-QA}. Inspired by the methods proposed in GenQA \cite{chen2024genqageneratingmillionsinstructions}, in this approach the model input is identical to the RAG approach, but is tasked to produce a list of 11 predictions (the same shape as the system prompt examples) instead of a single one. The output of the model is then a uniform random choice between the 11 predictions. Audio perception is not deterministic; the same prompt can yield multiple valid interpretations across different listeners. RAG-QA enhances the retrieval-based approach by generating multiple plausible recommendations, introducing a form of stochasticity that better reflects real-world subjectivity. This method aims to prevent collapse to a single interpretation and enables a broader exploration of the population preference space.
\end{itemize}

Figure \ref{fig:RAG} provides more details on the functionality of RAG and RAG-QA.

\subsection{Parameter Efficient Fine-Tuning}
In addition to ICL, we explore a few PEFT approaches. Instead of retraining all the parameters of a model, PEFT focuses on modifying only a small subset of parameters or tune the model through efficient reparameterization. This approach significantly reduces computational and storage costs while maintaining or even improving task-specific performance, compared to conventional fine-tuning \cite{fu2022effectivenessparameterefficientfinetuning,ding2023parameter,hu2022lora}. 

There are dozens of variations proposed \cite{houlsby2019parameterefficienttransferlearningnlp,pfeiffer2020adapterhubframeworkadaptingtransformers,radford2019language,lester2021powerscaleparameterefficientprompt,donahue2013decafdeepconvolutionalactivation,gheini2021crossattentionneedadaptingpretrained,lialin2024scalingscaleupguide} but for the purposes of this work we focus only on two. Prefix-Tuning \cite{li2021prefixtuningoptimizingcontinuousprompts} and Low Rank Adaptation (LoRA)\cite{hu2022lora}. The training for the following experiments is conducted on a single NVIDIA GeForce RTX 3090 with 24 GB of GPU memory.

\begin{figure*}
  \centering
  \includegraphics[width=\textwidth]{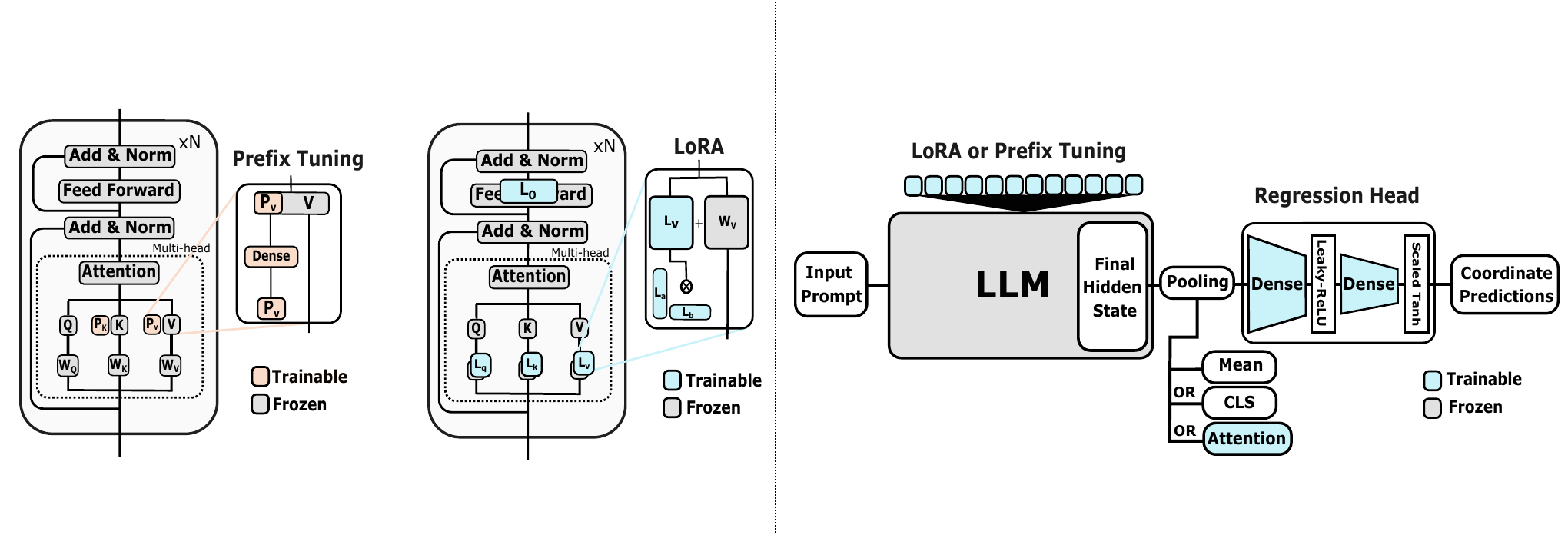}
  \caption{Parameter Efficient Fine-Tuning with regression head. The figure on the left illustrates the details of Prefix Tuning and LoRA methods, while the figure on the right showcases how the final hidden state of the LLM is mapped to Beosonic coordinates.}
  \label{fig:PEFT}
\end{figure*}

\begin{itemize}
\item \textbf{Prefix-Tuning} is a PEFT approach that optimizes a small continuous task-specific prefix, allowing subsequent tokens to attend to it as if it were a set of virtual ``tokens". The prefix activations are drawn from a trainable matrix \(P_\theta\) which is reparameterized by representing each row of \(P_\theta\) as the output of a feedforward neural network applied to a smaller matrix \(P'_\theta\). 

\item \textbf{LoRA} injects trainable rank decomposition matrices into each layer of the Transformer architecture, greatly reducing the number of trainable parameters for downstream tasks. The details of the two methods can be observed in Figure \ref{fig:PEFT} (left).
\end{itemize}

Our first method for optimizing the LLM involves pooling the final hidden state and passing it through a regression head. We explore several pooling techniques.
\begin{itemize}
    \item \textbf{Mean pooling:} Computes the average across the sequence length, condensing the representation to a fixed size.
    \item \textbf{CLS pooling:} Uses the special classification token (CLS) from the sequence as the representation.
    \item \textbf{Attention pooling:} Passes the final hidden state through a trainable attention layer to produce the representation. 
\end{itemize}
The underlying idea is to leverage the LLM's language understanding to produce a fine-tuned representation of the input prompt, which the regression head then maps to the Beosonic space. Figure \ref{fig:PEFT} (right) illustrates this approach.

The output of the regression head is an array of size $11 \times 2$, representing 11 distinct coordinate pairs in the Beosonic space. By training the model to predict this set of points simultaneously, we effectively task the model with predicting the empirical distribution of listener preferences for a given prompt. The loss function for optimizing the model is computed using the Sinkhorn divergence which is an affordable yet positive and definite approximation of the Optimal Transport (Wasserstein) distance \cite{feydy2018interpolatingoptimaltransportmmd}. This allows the model to learn the spread and clustering of human opinions.

The models are trained for 200 epochs, and we search for the hyperparameters that achieve the lowest validation loss. The hyperparameter pool includes all the combinations of Prefix Tuning with different numbers of virtual tokens, LoRA with different numbers of lower ranks, and whether or not data augmentation (see \ref{subsec:augmentation}) is utilized.

In addition to the regression approach, we use LoRA for a standard next-token prediction task. Here, the model generates a text string representing the full array of $11 \times 2$ coordinates. Similar to the regression approach, this trains the model to output a distribution of settings. 

To control the distributional spread during inference, one can adjust the LLM’s sampling temperature. For models that generate a single coordinate per forward pass, lowering the temperature collapses the predictions toward the most likely text sequence, effectively trading diversity for a ``safe" consistent point. However, for models that generate a full list of coordinates at once (such as the RAG-QA and PEFT approaches), the temperature does not govern the spatial spread of the points within the list; a low temperature simply ensures the model deterministically outputs the same diverse list every time. It should also be noted that across all models, excessively high temperatures risk degrading the required text formatting, such as the brackets used for coordinate arrays.

During inference, we decouple the training objective from the evaluation strategy. Regardless of whether a model outputs a single point or an array, we sample a single 2D coordinate per forward pass. To generate the full predictive distribution required for the evaluation metric, we execute multiple model calls and aggregate the sampled coordinates.

\subsection{Data Augmentation} \label{subsec:augmentation}
The 60 entries in the training set are more than enough to provide few-shot examples, but for PEFT this number might not be sufficient. For this reason, we experiment with data augmentation in the training set. We do this by using a custom dictionary of synonyms and generating different combinations of the same prompts, without changing the contextual meaning. Moreover, in an attempt to avoid overfitting, we add a sufficient amount of noise to the labels of the variants, to blur the targets while negligibly affecting them perceptually (less than 0.3~dB change in the equalization gains). We then test the performance of the PEFT models both with and without augmentation/label blurring, on the same, \emph{unedited} validation set. The blurring contributes to improved generalizability. With this approach we increase the size of our \emph{training} data by approximately 50 times.

\section{Evaluation Metrics} \label{sec:Evaluation}

As we have pointed out, all participants provide feedback for the same entries. This approach results in a dataset with only 120 prompts, whereas if each participant was labeling different entries, we would be able to prepare 1,320 unique prompt-audio pairs. However, we argue that the task the recommender system aims to solve is inherently subjective. Each individual may have different interpretations of the prompt, making single-point responses inadequate for capturing the full range of perceptual variation. 

\begin{figure}
\begin{center} 
    \includegraphics[width = \columnwidth]{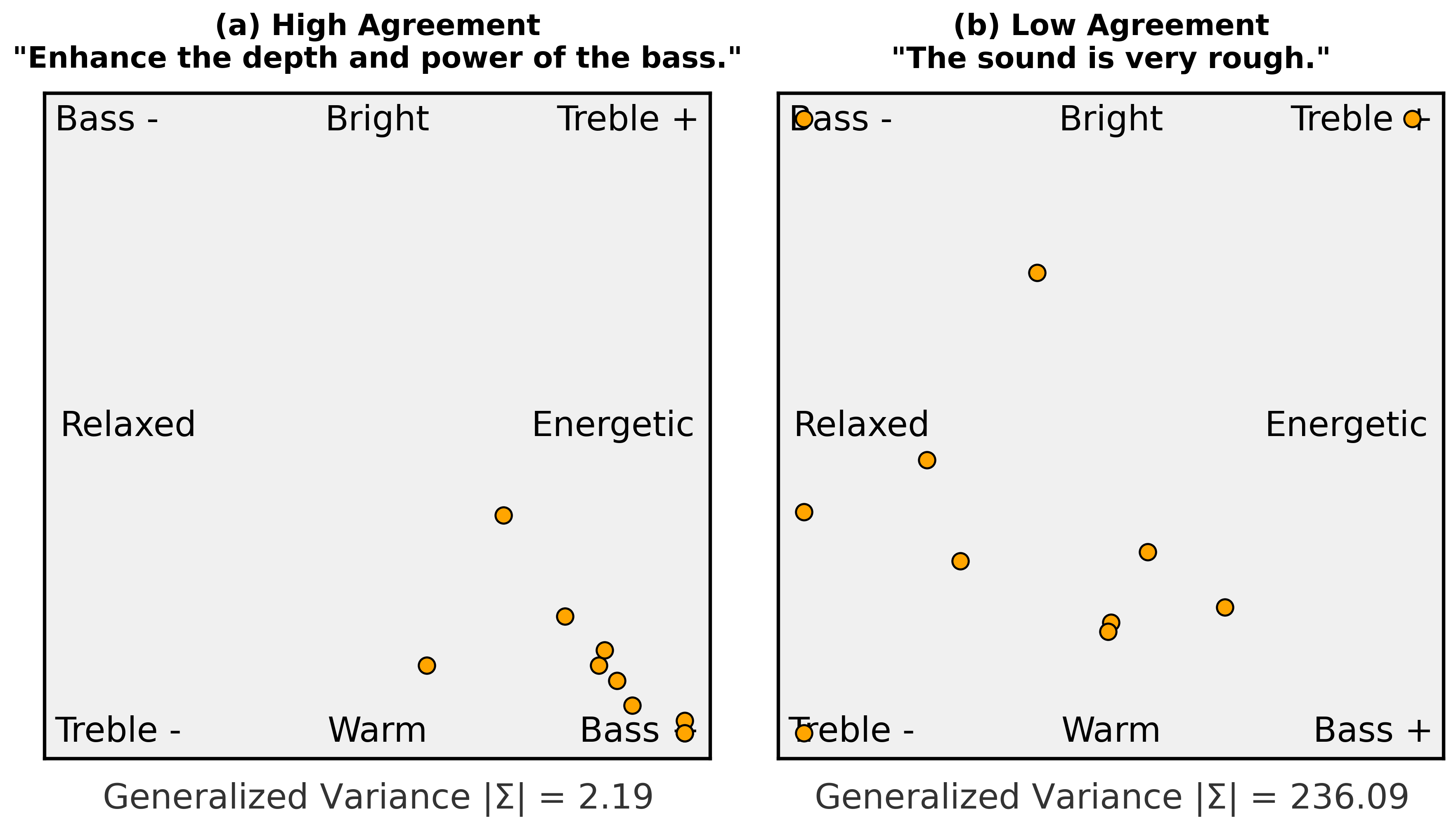}\ 
    \caption{Visualization of listener agreement in the Beosonic space. (a) For specific prompts, users agree more (low Generalized Variance $|\Sigma|$), indicating a clear consensus. (b) For ambiguous prompts, preferences are widely dispersed (high $|\Sigma|$), demonstrating that a single ``correct" equalization setting does not exist.} 
    \label{fig:responses} 
\end{center} 
\end{figure}

Figure \ref{fig:responses} illustrates with an example why this approach is superior to collecting only single-point responses. In case (b), no single label can fully represent the appropriate response, as it fails to capture the broader distribution of listener responses. The variation is intuitive: When given a vague prompt such as \emph{``The sound is very rough",} a natural human reaction would be to request further explanation. Moreover, what one person perceives as rough may not sound rough to someone else, reinforcing the idea that perception varies widely across individuals.

Thus, rather than seeking a single ``correct" frequency response, the key takeaway is that no universally ideal frequency response exists---only a distribution of plausible responses reflecting different listener perceptions.
 
This example highlights three important considerations.
\begin{itemize}
    \item The evaluation of the artificial equalizers should account for the inherent subjectivity in both audio and language perception.
    \item An ideal recommender system should incorporate personalization capabilities to adapt to individual listener preferences.
    \item In a real-world setting, a lot of prompts cannot directly be achieved through frequency response adjustments alone. For this reason, there is significant potential in exploring additional loudspeaker parameters (e.g. reverberation \cite{doh2025can, chu2025text2fx} for commands such as ``Make it sound like a large church") for better alignment to different user requirements.
\end{itemize}

The evaluation metrics that follow were designed with these three considerations in mind.

\subsection{Kantorovich Distance}
Every prompt in the data is represented by 11 coordinates in the Beosonic space, each corresponding to a response from a participant. In order to evaluate how well a model can map natural language to the Beosonic space, we employ the Kantorovich-1 distance \cite{kantorovich1960mathematical} (more commonly \cite{vershik2013long} known as the Wasserstein or Earth Mover’s Distance), to compare the predicted distribution with the ``true" distribution of the responses.

We argue that this metric is well suited for our task because unlike point-wise error metrics, such as using a centroid or calculating the Mean Squared Error (MSE), the Kantorovich distance captures differences in the overall shape, spread, and central tendencies of distributions. This can be exemplified with the case of a bimodal distribution: If the ``true" distribution has two distant modes, then minimizing MSE or computing the centroid leads to averaging effects---placing predictions between the modes---where there may be no meaningful probability mass.

Furthermore, the use of the Kullback-Leibler divergence may not be suited either, since participant responses appear to cluster along specific trajectories in the Beosonic space, meaning that they lie on a lower-dimensional manifold rather than being spread out across the full space \cite{arjovsky2017wassersteingan}.

Another candidate for quantifying the similarity between the collected and the generated samples in the 2D space is the Fasano-Franceschini test \cite{fasano1987multidimensional}. This test extends the Kolmogorov-Smirnov test to multiple dimensions, but it lacks the convenient optimization properties provided by the Kantorovich distance, and is thus less suited for our applications.

\begin{figure*}[!t]
\begin{center} 
    \includegraphics[width=\textwidth]{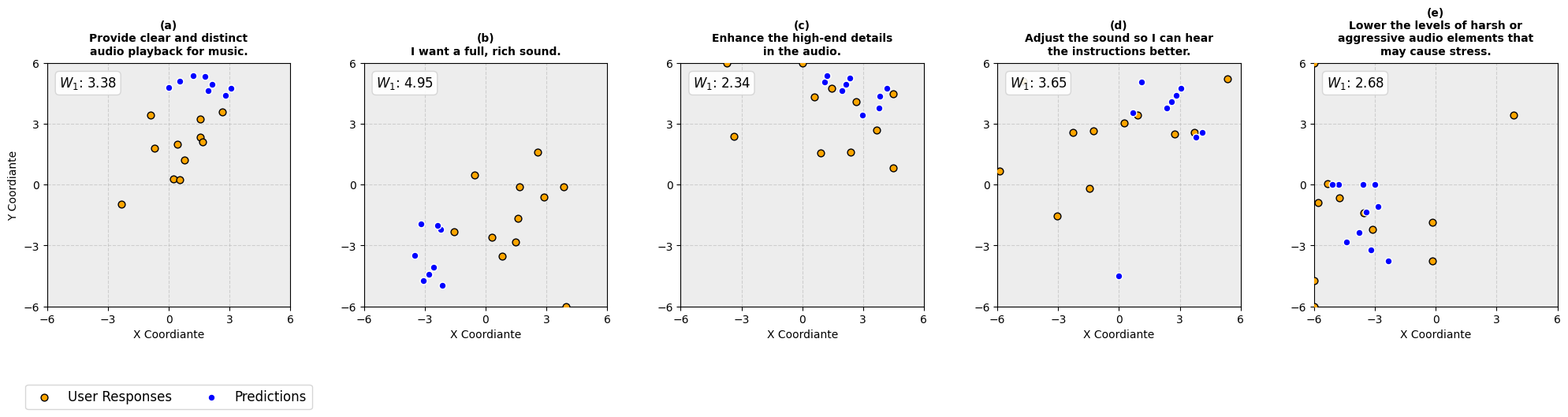}
    \caption{\textbf{Visualization of the Kantorovich ($W_1$) distance across varying degrees of listener disagreement.} The plots display five prompts, ordered from lowest ground-truth variance to highest variance. The human responses are shown in orange alongside the predictive distribution of text2beosonic in blue.}
    \label{fig:beosonic_pred_5_exp} 
\end{center} 
\end{figure*}

\begin{figure}
\begin{center} 
    \includegraphics[width=\columnwidth]{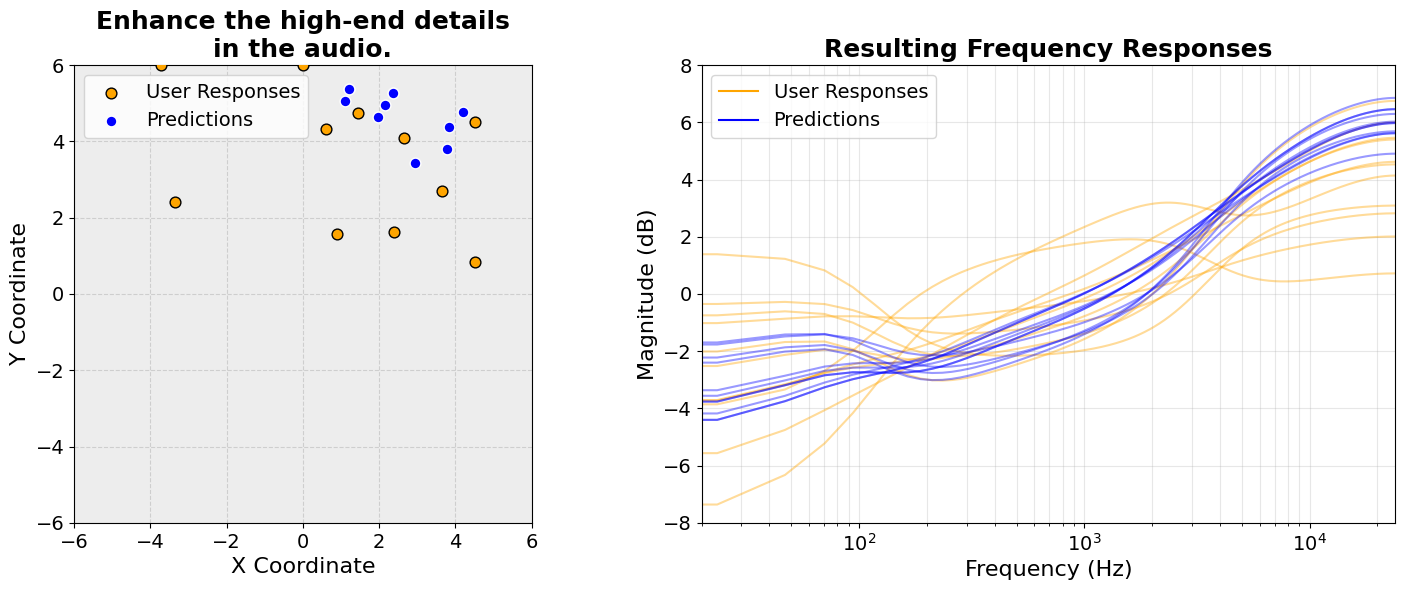}
    \caption{\textbf{Visualization of the user responses and model predictions in the transfer function space.} As the prompt suggests, we can observe a boost in the high frequencies and some attenuation in the lows, for both samples.}
    \label{fig:transfer_funcs} 
\end{center} 
\end{figure}

Mathematically, the Kantorovich-1 distance between two probability distributions \(\mu\) and \(\nu\) is defined as:

\begin{equation}
\label{wass}
W_1(\mu, \nu) = \inf_{\gamma \in \Pi(\mu, \nu)} \mathbb{E}_{(x,y) \sim \gamma} \left[ \|x - y\|_2 \right],
\end{equation}

where \(\Pi(\mu, \nu)\) denotes the set of all joint distributions (couplings) with marginals \(\mu\) and \(\nu\). In practice, the distance is usually calculated using Linear Programming (LP) and to circumvent the approximation of \(\Pi(\mu, \nu)\), the LP problem is formulated in its dual form. This formulation allows us to quantify the discrepancy between the predictive distribution and the actual distribution of listener responses in the Beosonic space. A few examples of the prompts, the responses of the participants, the predictions of the text2beosonic baseline and the $W_1$ value are presented in Figure \ref{fig:beosonic_pred_5_exp}. We also illustrate the distribution of the transfer functions for one of the examples in Figure \ref{fig:transfer_funcs}.

Furthermore, the sampling approach enables seamless integration of Reinforcement Learning from Human Feedback (RLHF) techniques, such as Direct Preference Optimization (DPO) \cite{rafailov2023direct}, by treating explicit or implicit feedback on each sample as a preference entry, thereby enabling effective personalization.

Finally, the proposed approach naturally extends to additional acoustic parameters, since it is applicable to any parameterization and number of parameters, with the Kantorovich/Wasserstein distance being tractable in arbitrarily high dimensions via its sliced variants \cite{deshpande2018generative}.

\subsection{Reflective Kantorovich Distance}
Since the Beosonic space is a bounded domain, standard Kernel Density Estimation (KDE) suffers from significant \textit{boundary bias}. This is a well-documented statistical phenomenon where probability mass is smoothed into the invalid domain, leading to a systematic underestimation of density at the boundaries \cite{jones1993simple, silverman2018density}. 

\begin{figure*}
    \includegraphics[width = \textwidth]{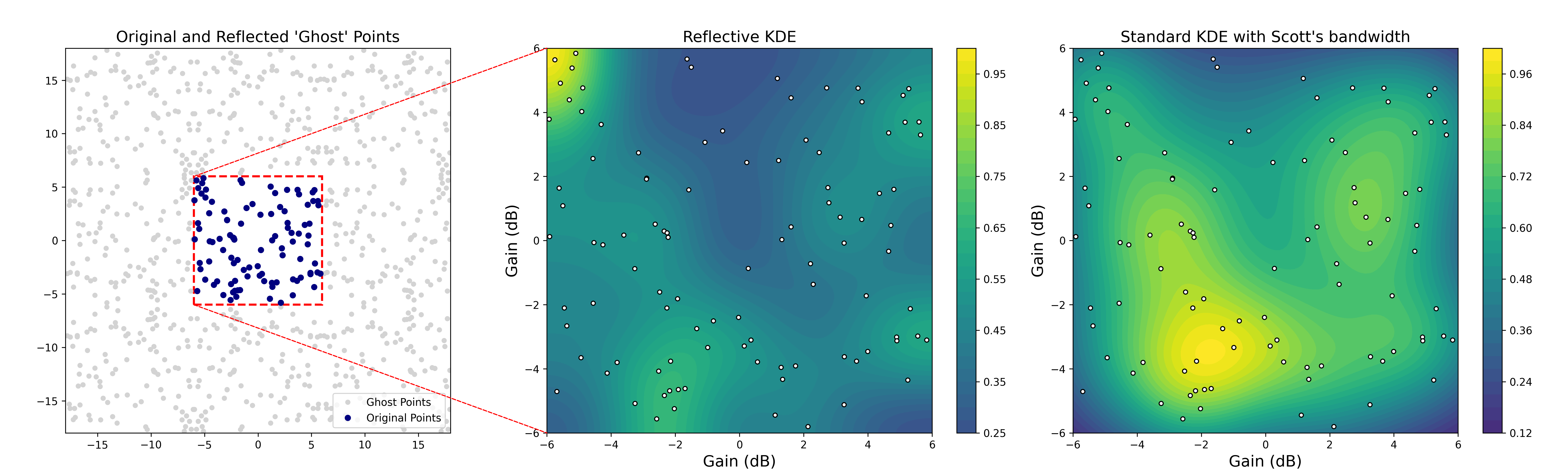}\ 
    \caption{Example of standard KDE vs reflective KDE. The blue/white points are drawn from a uniform distribution in the [-6,6]x[-6,6] square. With the reflection heuristic the density range is restricted, resulting in a better approximation of the uniform distribution. Border underestimation is also reduced.} 
    \label{fig:kde} 
\end{figure*}

This bias is particularly detrimental in the context of equalizer control, where user preferences frequently cluster at the extrema of the interface (e.g., selecting ``maximum bass"). A standard KDE would artificially dampen the probability density of these valid choices, subsequently distorting any distance metric derived from it. To resolve this, we employ reflective KDE \cite{schuster1985incorporating}, a boundary correction technique that mirrors data points across the edges of the support. This transforms the density estimator from a biased estimator (near boundaries) to a consistent one, ensuring that the ``hard stops" of the physical controller are accurately reflected in the preference distribution (Figure \ref{fig:kde}).

An important parameter in KDE is the bandwidth of the kernel, which controls the smoothness of the estimated density. The suboptimal estimation shown in Figure \ref{fig:kde} can likely be attributed to using too small a bandwidth. To investigate this, we experiment with different bandwidth values on simulated data, using Scott’s rule \cite{scott2010scott} as a starting point. We find that, in addition to yielding improved density estimates, the use of the reflection heuristic significantly increases robustness to bandwidth variation.

\begin{figure*}
    \includegraphics[width = \textwidth]{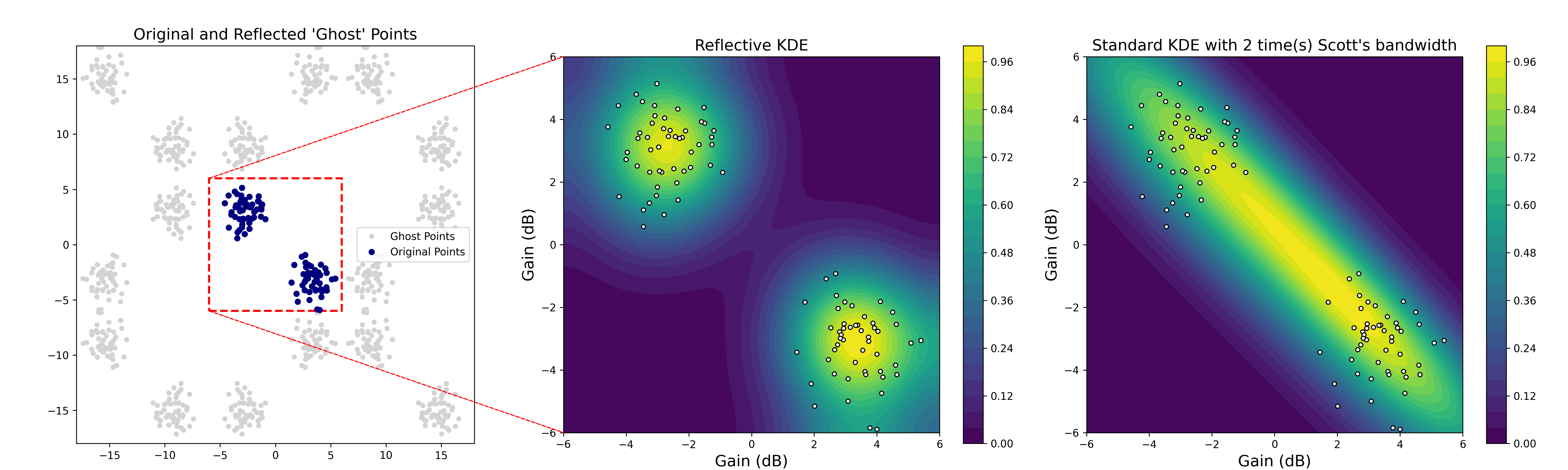}\ 
    \caption{Example of standard KDE vs reflective KDE. The blue/white points are drawn from a bimodal standard Gaussian distribution with modes [-3,3] and [3,-3]. With the reflection heuristic the density estimation separates between the two modes, resulting in a better approximation of the distribution. Standard KDE fails to separate between modes, especially at larger bandwidths.} 
    \label{fig:kde2} 
\end{figure*}

Figure \ref{fig:kde2} illustrates that simply increasing the KDE bandwidth does not adequately resolve the boundary underestimation problem. In fact, in the bimodal case, a larger bandwidth causes the two modes to merge, failing to reflect the true shape of the distribution. In contrast, reflective KDE accurately captures both modes.

This improvement is particularly important when the underlying distribution is defined on a bounded domain. In such cases, the relative distance between two modes becomes more meaningful than in unbounded (e.g., infinite) domains, where large-scale context diminishes the perceived separation. Within a bounded support, the two point clouds---each generated from a distinct mode of a bimodal Gaussian---are sufficiently far apart to warrant being treated as separate modes. Reflective KDE preserves this structure, while standard KDE may blur it away. 

Finally, we also considered a Bayesian approach that incorporates a scaled 2D Beta variant prior, accounting for our limited collected data. However, this parametric method was deemed inadequate in capturing the complex variations present within the Beosonic space. 

Consequently, we propose the usage of the \textbf{Reflective Kantorovich Distance}. It is important to clarify that this does not constitute a new distance metric; mathematically, it remains the standard Kantorovich distance. The distinction lies in the preprocessing: by calculating the distance over densities estimated via reflective KDE, we ensure that the metric measures divergence from a boundary-aware representation of population preference.

\subsection{Metric Interpretability}
While the Kantorovich distance is formally a statistical measure of mass transport, it provides a crucial intuition regarding the perceptual validity of the generated audio settings. In our context, this metric can be interpreted as a measure of \textit{Plausibility Alignment}.

A low Kantorovich distance indicates that the model distributes its probability mass exactly where the human population distributes their preferences. This manifests in two distinct ways depending on the nature of the prompt:
\begin{itemize}
    \item For \textbf{specific prompts} where human agreement is high (low variance), a low distance implies the model reliably reproduces the distinct sound character requested by the consensus.
    \item For \textbf{ambiguous prompts} where human agreement is low (high variance), a low distance implies the model successfully captures the diversity of valid interpretations. It avoids collapsing to a ``safe'' average (which could fail to fully satisfy any specific user) and avoids hallucinating settings that no human listener would choose.
\end{itemize}

\begin{figure}
\begin{center} 
    \includegraphics[width = \columnwidth]{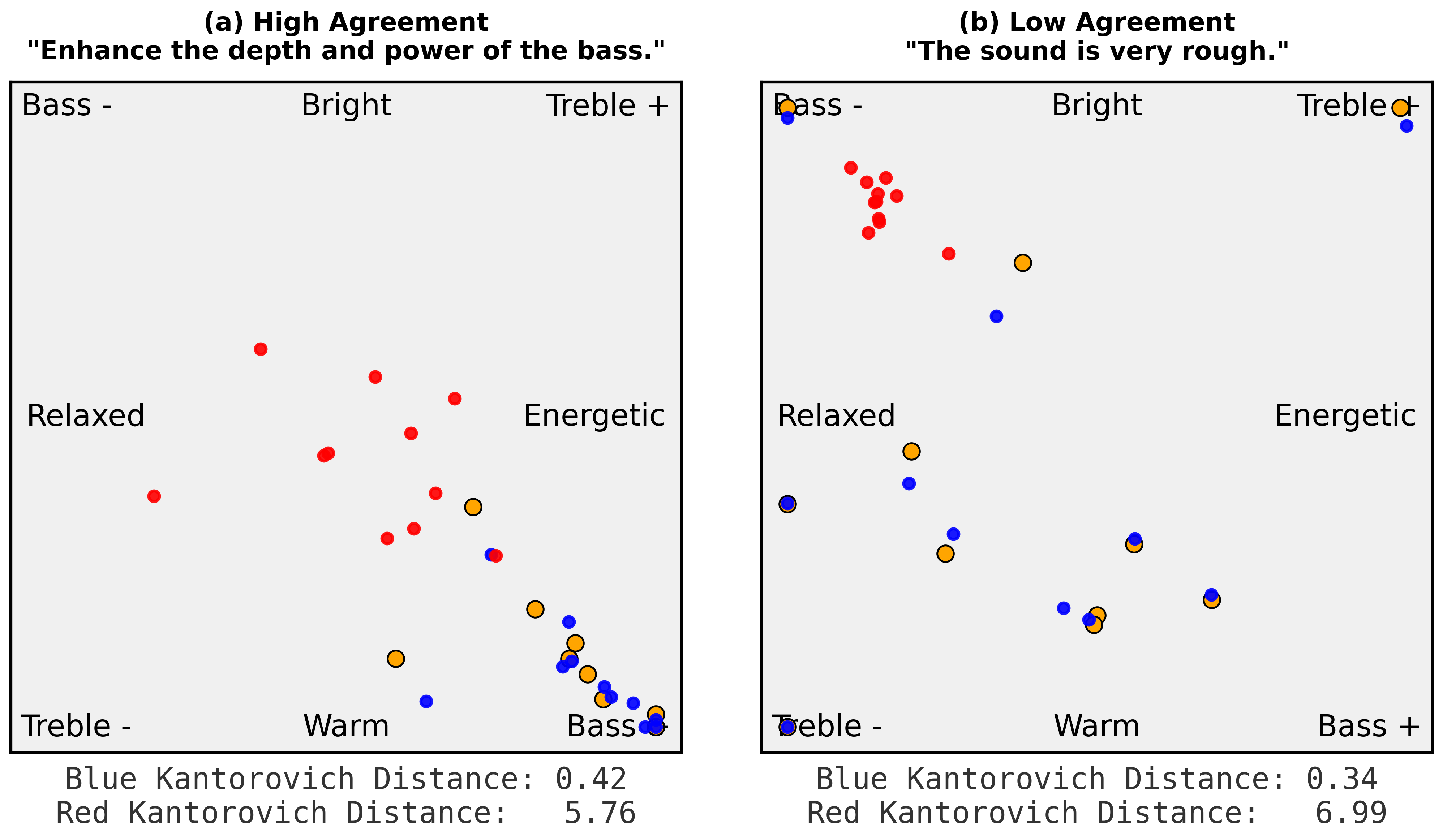}\ 
    \caption{\textbf{Visual validation of the Kantorovich distance as a performance metric.} 
    The plots contrast the Ground Truth user responses (Orange) against two synthetic model predictions: a ``Population-Aligned'' distribution (Blue) and a ``Misaligned'' distribution (Red). 
    \textbf{(a) Specific Prompt:} The Red samples are spatially shifted from the consensus. The Kantorovich distance accurately reflects this error with a high value compared to the matching Blue samples.
    \textbf{(b) Ambiguous Prompt:} The user preferences are widely dispersed. The Red samples simulate a deterministic model that collapses to a ``safe'' average. However, the Kantorovich distance correctly penalizes this lack of diversity, while rewarding the Blue samples for successfully capturing the spread of the population's preferences.}
    \label{fig:responses_kantorovich_v2} 
\end{center} 
\end{figure}

Conversely, a high Kantorovich distance signifies a fundamental semantic disconnect: the model is suggesting equalization settings that lie outside the ``zone of plausibility'' defined by human listeners (Figure \ref{fig:responses_kantorovich_v2}). Thus, the metric intuitively quantifies the model's ability to avoid ``illogical'' or ``unnatural'' mappings, ensuring that every generated setting--whether consistent or varied--aligns with a human perception of the prompt.

\section{Results \& Discussion} \label{sec:Results}

Our evaluation examines the performance of both ICL and PEFT approaches using the objective metrics described in Section \ref{sec:Evaluation}. For every method, predictions are generated for all entries in the test set. 

\subsection{Base LLM Comparison}

\begin{figure}
\begin{center} 
    \includegraphics[width=0.85\linewidth]{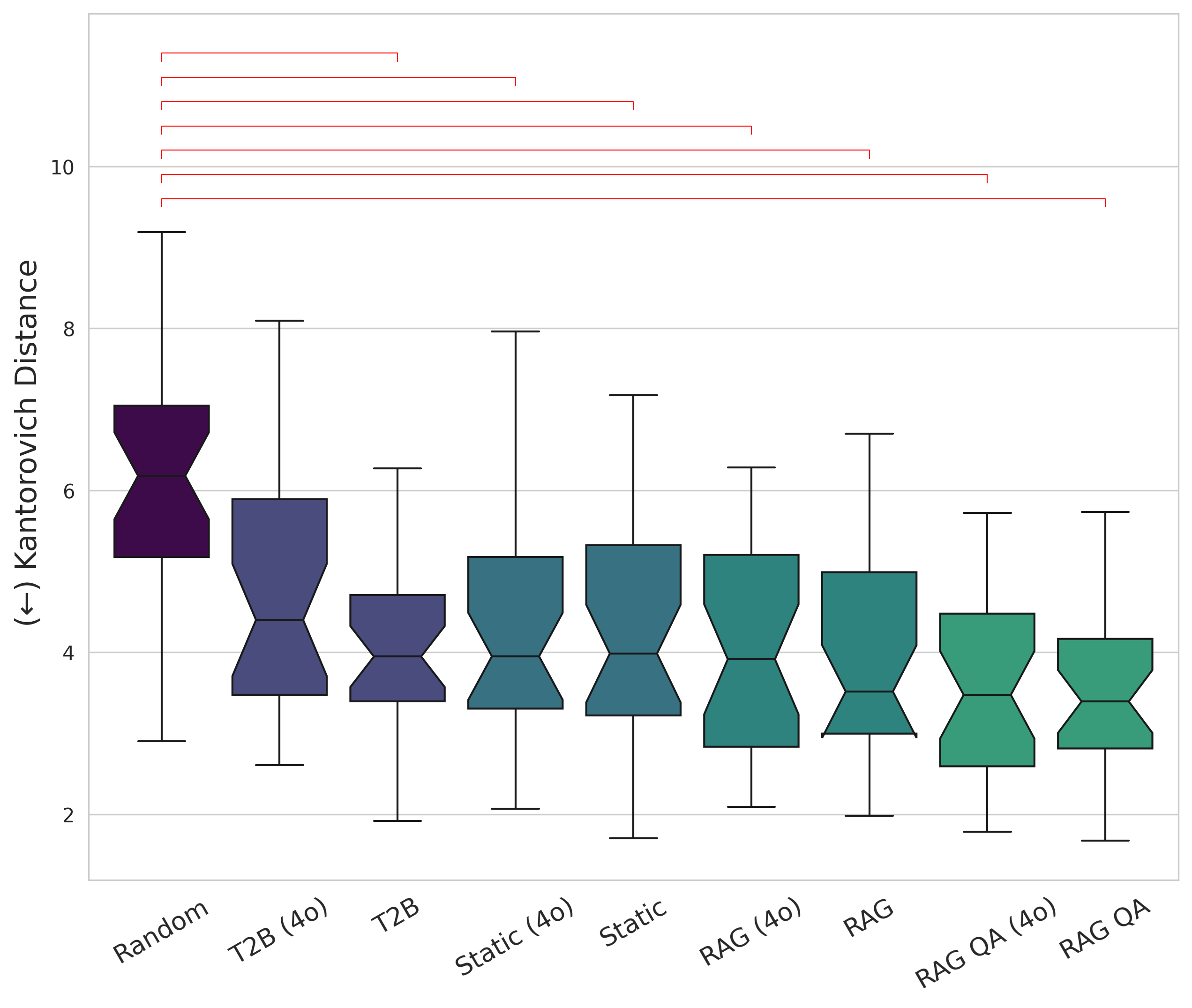}\ 
    \caption{Comparison between GPT-4o mini and Phi-3.5 mini on the proposed ICL methods using the Kantorovich Distance. Horizontal brackets indicate significant difference between distributions at p $\leq$ 0.05 (black) or p $\leq$ 0.01 (red) levels. In this case all p-values are less than 0.01 or over 0.05 so no black brackets appear.} 
    \label{fig:wass_icl} 
\end{center} 
\end{figure}

\begin{figure}
\begin{center} 
    \includegraphics[width=0.85\linewidth]{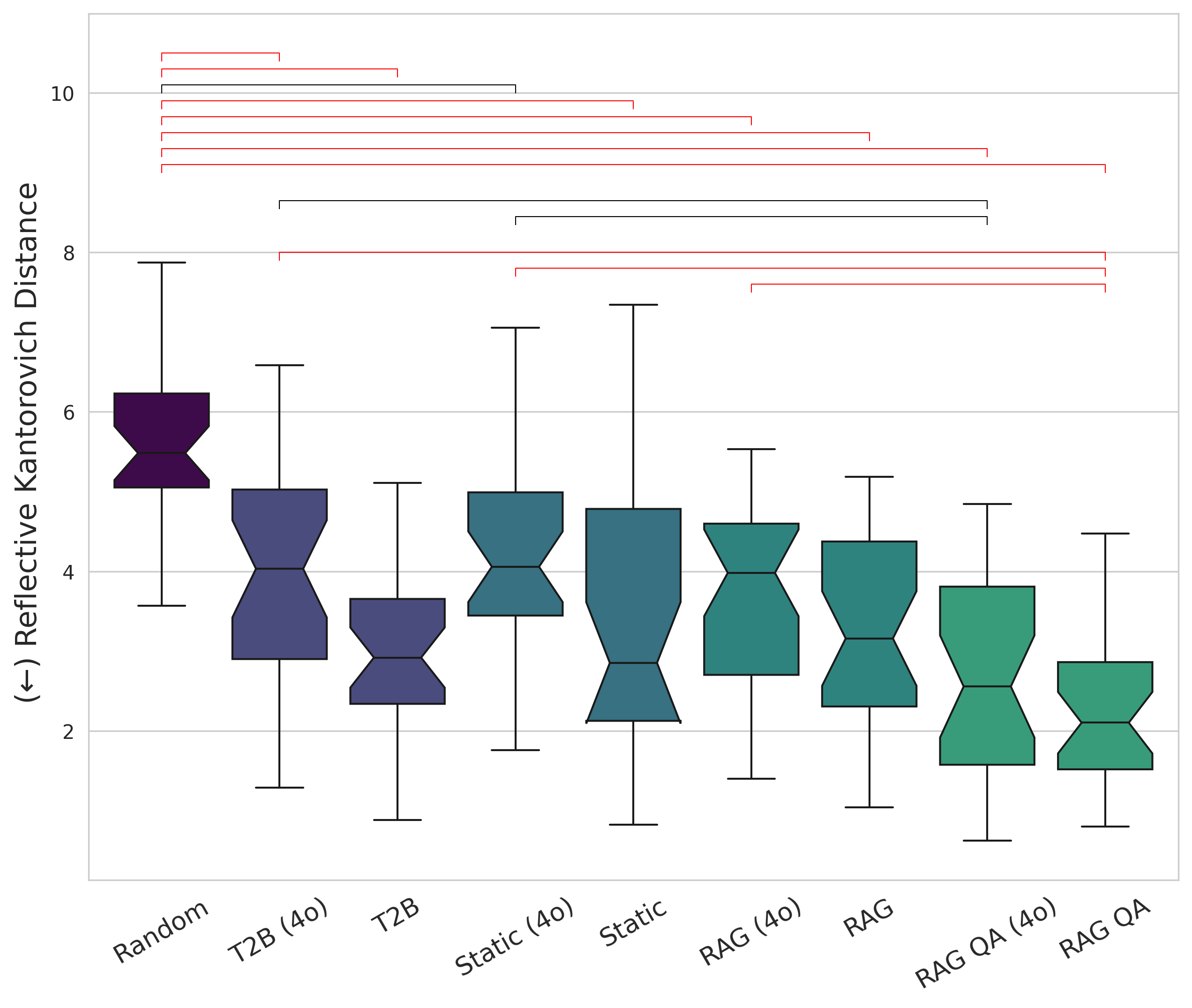}\ 
    \caption{Comparison between GPT-4o mini and Phi-3.5 mini on the proposed ICL methods using the reflective Kantorovich Distance. Horizontal brackets indicate significant difference between distributions at p $\leq$ 0.05 (black) or p $\leq$ 0.01 (red) levels.} 
    \label{fig:rwass_icl} 
\end{center} 
\end{figure}

For ICL methods specifically, we use Phi-3.5 mini but also assess GPT-4o mini as the underlying large language model. Figures \ref{fig:wass_icl} and \ref{fig:rwass_icl} present a comparison between the predicted and ground-truth response distributions, evaluated using the Kantorovich and reflective Kantorovich distances, respectively. Additionally, notched boxplots \cite{mcgill1978variations} are employed to approximate the 95\% confidence intervals around the medians.

We evaluate differences among model performance through a series of statistical tests. First, a Kruskal–Wallis test \cite{kruskal1952use} is performed to determine whether the distributions of the metrics differ among the models. A significant result (p $\leq$ 0.05) indicates that at least one group deviates from the others. Next, to identify which specific groups differ, a Dunn’s post hoc test \cite{dunn1964multiple} is applied with Bonferroni correction to adjust for multiple comparisons. Significant pairwise differences are visually annotated, with brackets indicating if two distributions are significantly different. We compare all methods with samples from randomly generated bi-variate Gaussians, to simulate random guessing.

Despite its smaller size, Phi-3.5 mini consistently yields lower median distance than GPT-4o mini, which could in part be attributed to its higher long-context performance. However, this difference is not statistically significant. Conversely, all methods perform significantly better than random guessing, for both metrics.

\subsection{Training Approach Comparison}
\begin{figure}
\begin{center} 
    \includegraphics[width=0.85\linewidth]{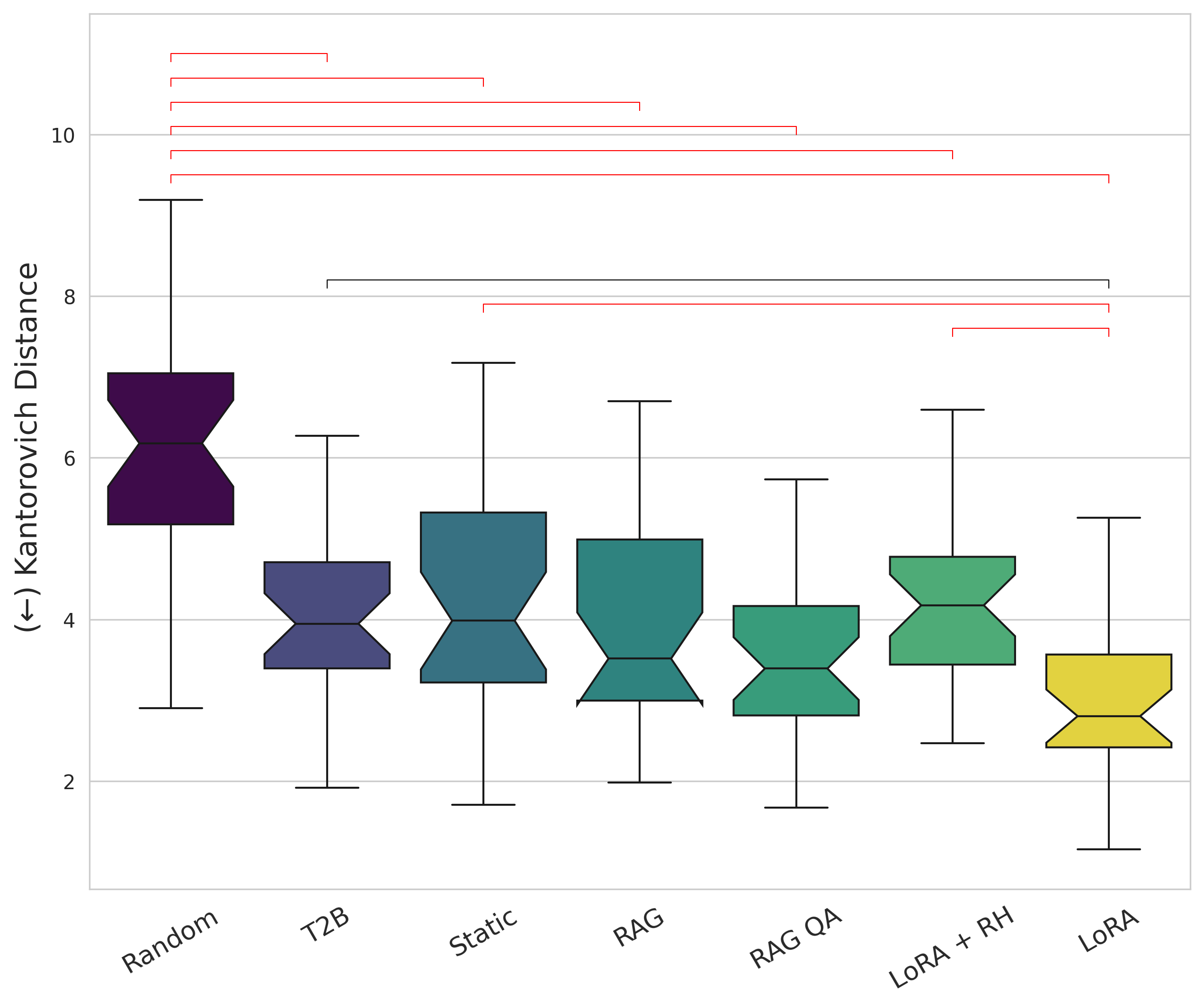}\ 
    \caption{Comparison between proposed ICL and PEFT methods using the Kantorovich Distance. Horizontal brackets indicate significant difference between distributions at p $\leq$ 0.05 (black) or p $\leq$ 0.01 (red) levels.} 
    \label{fig:wass_peft} 
\end{center} 
\end{figure}

\begin{figure}
\begin{center} 
    \includegraphics[width=0.85\linewidth]{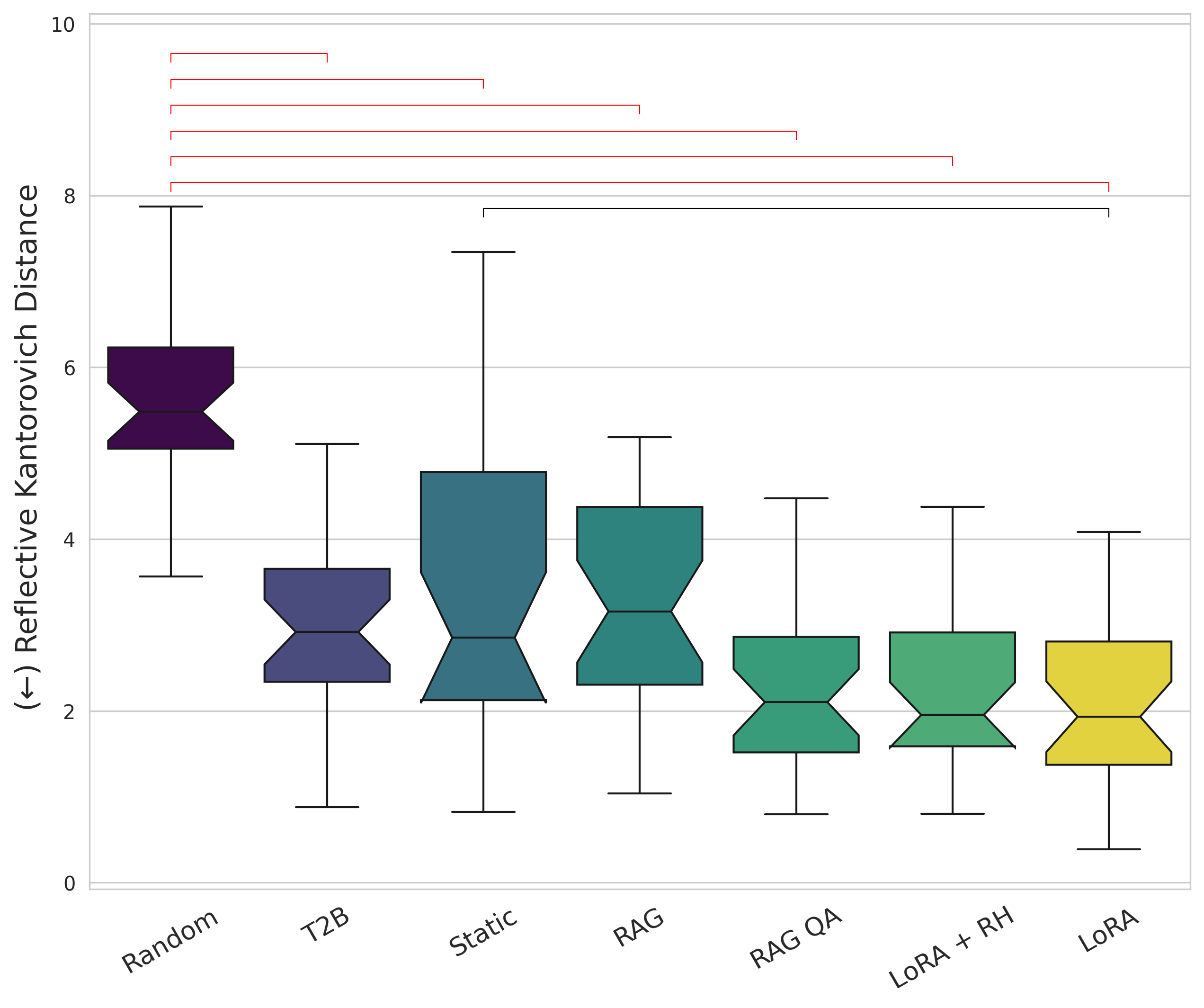}\ 
    \caption{Comparison between proposed ICL and PEFT methods using the reflective Kantorovich Distance. Horizontal brackets indicate significant difference between distributions at p $\leq$ 0.05 (black) or p $\leq$ 0.01 (red) levels.} 
    \label{fig:rwass_peft} 
\end{center} 
\end{figure}

Subsequently, we benchmark the best performing ICL configurations against our top PEFT approaches, with Phi-3.5 mini as the base model. As illustrated in Figures \ref{fig:wass_peft} and \ref{fig:rwass_peft}, leveraging LoRA for next-token-prediction outperforms its optimized regression head counterpart for the Kantorovich distance, and achieves the lowest median distance for both metrics. However, the benefits of fine-tuning on limited but carefully augmented data, do not suffice to provide a statistically significant improvement over the ICL methods. We speculate that 11 responses may not be enough to capture the distribution of the overall population and more extensive data collection experiments could play a critical role in performance improvements. 

Overall, our results indicate that both ICL and PEFT can translate natural‑language prompts into plausible equalization settings, evaluated by distributional metrics that capture listening‑context effects more faithfully than single‑point errors.

\subsection{Limitations \& Future work}
The primary contribution of this work is to establish the feasibility of inclusive, LLM-based equalization and to propose a corresponding computational evaluation framework. However, we recognize three key limitations. 

First, our data volume is limited (120 prompts), constraining statistical power. The density of annotations (11 per prompt) allowed for a distributional approach, but is also of limited size. 

Secondly, this study utilized text-only conditioning. We deliberately excluded audio analysis to isolate the challenge of the semantic gap without the confounding variable of signal content. The decision to rely on text-only conditioning was also motivated by practical deployment constraints. Incorporating audio analysis would require the system to continuously process high-dimensional signal embeddings to contextually adapt to the content. This introduces substantial complexity and latency, which is often prohibitive for real-time applications. By decoupling the control logic from the signal path, our text-only framework enables a sparse inference, on-demand implementation that is far more suitable for integration into resource-constrained edge devices. However, audio conditioning (e.g., using CLAP embeddings or signal analysis) is essential for a fully context-aware system. 

Third, and most importantly, we have not yet conducted perceptual validation with human listeners. While we minimize the Kantorovich distance to human preference distributions, strictly speaking, this validates that the model mimics human choices, not necessarily that it optimizes satisfaction for a new user. This approach however, sets the stage for future work on improved audio reproduction through personalization. Incorporating additional loudspeaker parameters is also a promising avenue of future exploration. 

\section{Conclusion} \label{sec:Conclusion}
This study introduces a framework for refining audio reproduction through LLM-based equalization, to bridge the gap between natural language descriptions and plausible frequency response adjustments. Unlike prior approaches that treat equalization as a deterministic regression problem, we formulate it as a distributional modeling task, acknowledging that natural language expression and audio perception are inherently subjective.

By integrating LLMs with both in-context learning and parameter-efficient fine-tuning strategies, and evaluating them using Optimal Transport metrics (Kantorovich distance), we demonstrated that LLMs can reliably model the diversity of user intent. Our system does not simply guess a single ``correct" setting, but predicts a distribution of plausible settings that statistically aligns with human disagreement patterns.

These results underscore the feasibility of using LLMs to capture the subjective landscape of audio control. While future work must integrate audio signal analysis and perform perceptual validation to confirm user satisfaction, this work lays the necessary computational foundation: treating natural language not as a command for a single parameter, but as a window into a distribution of user preferences.

    \section*{Acknowledgments}
Special thanks to Francesc Lluís Salvadó for developing the original text2beosonic, Martin Bo Møller for his insightful suggestions as well as Geoff Martin and Jakob Dyreby for their support with Beosonic.

We would also like to thank our colleagues from Surrey University, Prof. Philip J. B. Jackson and Prof. Wenwu Wang for their valuable discussions under the AURIC project.

This work was jointly funded by Bang \& Olufsen A/S and The Innovation Fund Denmark.

\bibliographystyle{IEEEtran}
\bibliography{bibfile}

\newpage

\section{Biography Section}

\begin{IEEEbiography}[{\includegraphics[width=1in,height=1.25in,clip,keepaspectratio]{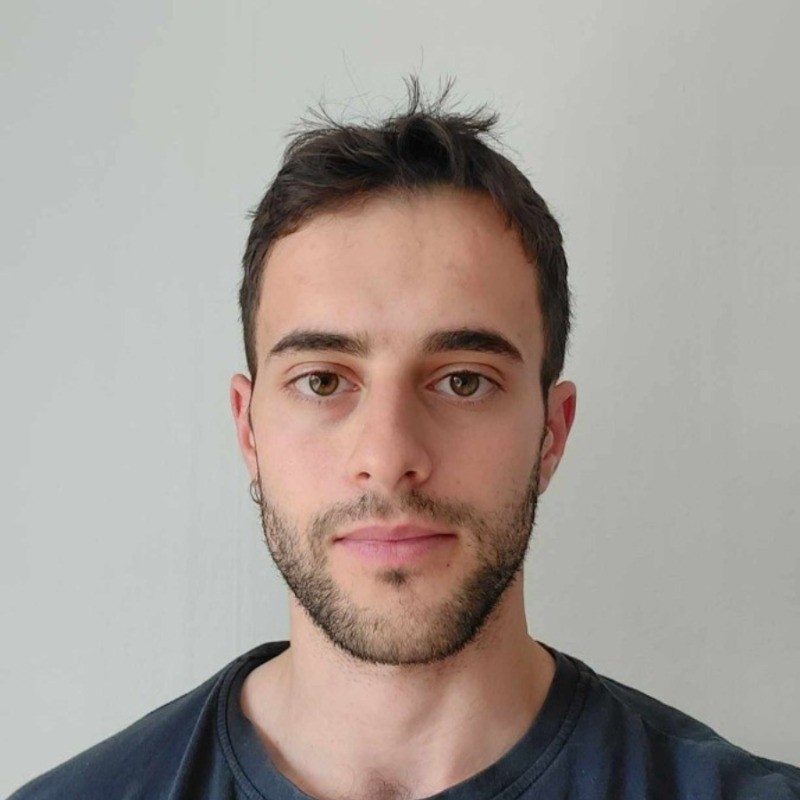}}]{Ioannis Stylianou} earned a BSc in Mathematics and Applied Mathematics from the University of Crete and an MSc in Statistics with Data Science from the University of Edinburgh. His initial research at the Foundation for Research and Technology - Hellas focused on implementing machine learning methods for protein engineering and folding. He subsequently joined Aalborg University as a Research Assistant, where he developed an open-source dataset for voice activity detection. Ioannis is currently an Industrial PhD candidate at Aalborg University in collaboration with Bang \& Olufsen A/S. His present research focuses on developing Large Language Model frameworks to map natural language prompts to complex loudspeaker parameters, utilizing reinforcement learning to account for subjective listener variability in audio control.
\end{IEEEbiography}

\begin{IEEEbiography}[{\includegraphics[width=1in,height=1.25in,clip,keepaspectratio]{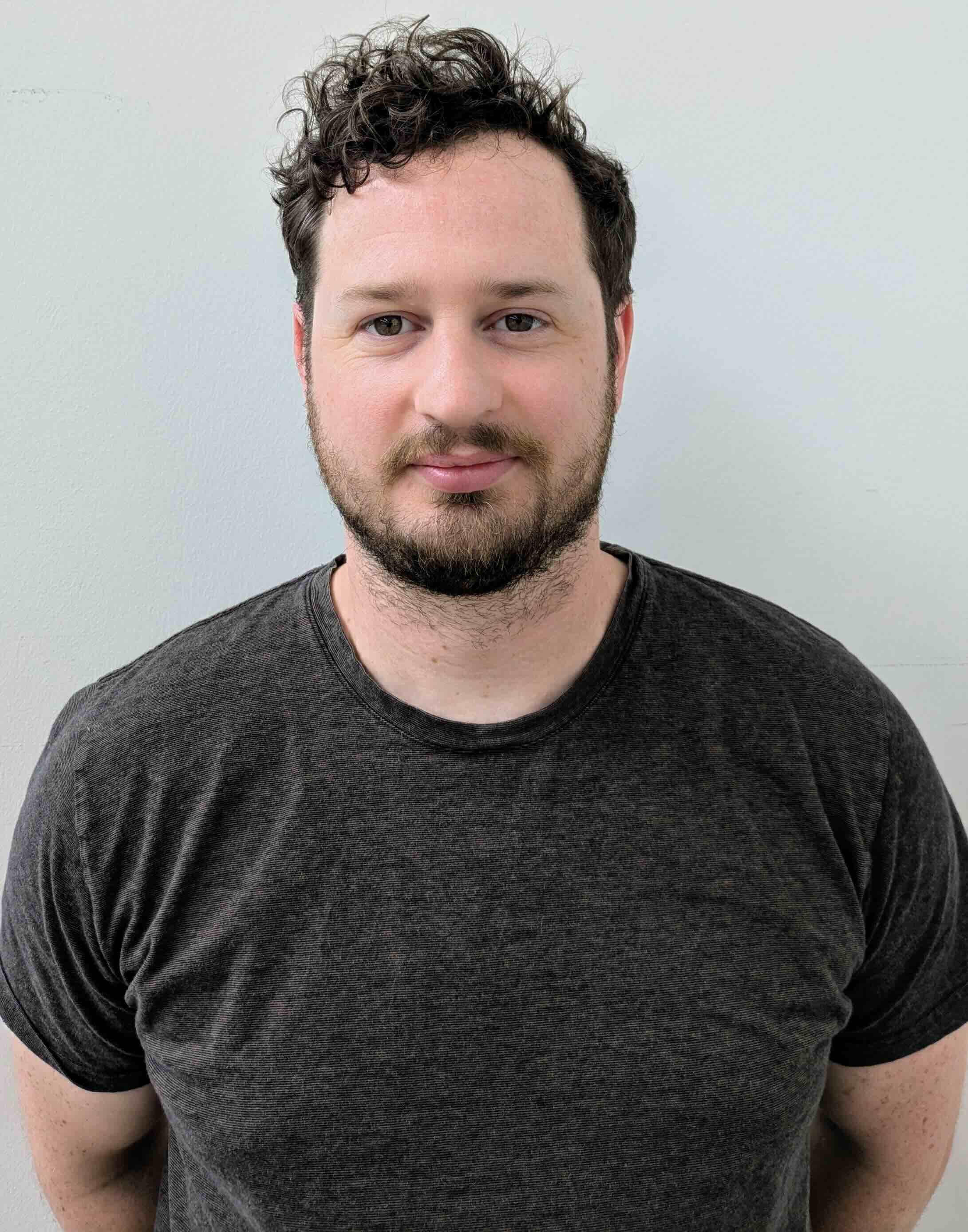}}]{Jon Francombe} graduated with a first-class honours degree in music and sound recording (Tonmeister) from the University of Surrey, Guildford, UK, in 2010, and received a Ph.D. in perceptual audio quality evaluation from the same institution in 2014. He then worked as a research fellow on the EPSRC-funded ``S3A: Future Spatial Audio for an Immersive Listener Experience at Home" project, investigating the perceptual attributes of spatial audio reproduction and new methods for immersive audio reproduction. This was followed by four years in the Audio Team at BBC Research \& Development, where he co-led the team and continued development of technology for delivering spatial audio experiences over connected personal devices. Jon currently works as a senior specialist in audio perception at Bang \& Olufsen, where he designs and conducts listening experiments and supervises various research projects.
\end{IEEEbiography}

\begin{IEEEbiography}[{\includegraphics[width=1in,height=1.25in,clip,keepaspectratio]{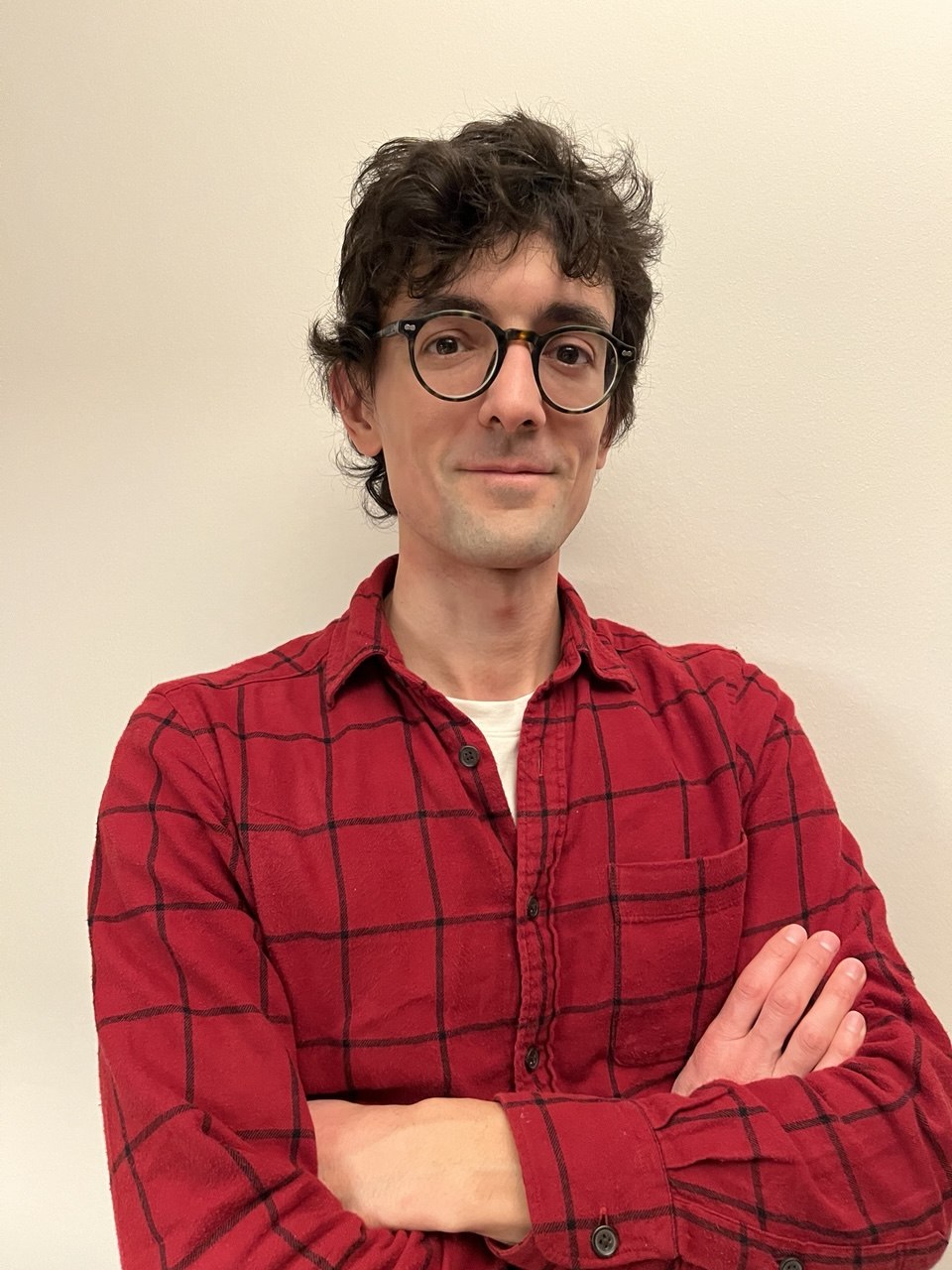}}]{Pablo Martínez-Nuevo} received the Ingeniero de Telecomunicación degree from the ETSIT, University of Valladolid, Spain, in 2010, and the M.Sc. degree in Electrical Engineering from Columbia University, NY, in 2012, where he was awarded the Edwin Howard Armstrong Memorial Award. He received the Ph.D. degree in Electrical Engineering and Computer Science from the Massachusetts Institute of Technology (MIT) in 2016 under the supervision of Prof. Alan V. Oppenheim in the Digital Signal Processing Group. He has been a recipient of the Fundación Rafael del Pino Fellowship, Madrid, Spain, for graduate studies at MIT. He joined Bang \& Olufsen, Denmark, in 2017, where he leads the company's artificial intelligence function and its academic collaborations. He is a co-inventor on several patents, and supervises industrial Ph.D. candidates in partnership with European universities. His research interests lie broadly within the areas of artificial intelligence, and signal and information processing.
\end{IEEEbiography}

\begin{IEEEbiography}[{\includegraphics[width=1in,height=1.25in,clip,keepaspectratio]{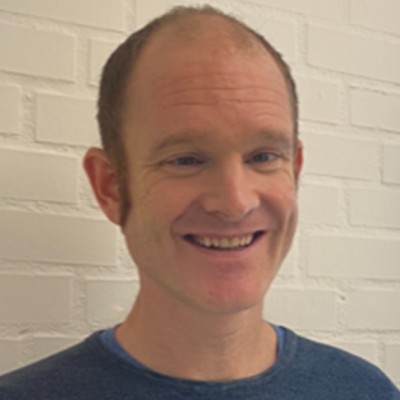}}]{Sven Ewan Shepstone} (Senior Member, IEEE) received the B.S. and M.S. degrees in Electrical Engineering from the University of Cape Town, South Africa, in 1999 and 2002, respectively. From 2005 to 2010, he was a Systems Engineer with Ericsson A/S, Denmark, where he worked on Ethernet and broadband communication systems. Since 2010, he has been with Bang \& Olufsen A/S, Denmark, where he is currently a Senior Specialist working on artificial intelligence technologies for consumer audio products. He received the Ph.D. degree from Aalborg University, Denmark, in 2015. His research interests include machine learning and artificial intelligence, with particular emphasis on deep learning, large language models, computer vision, speech and audio processing, diffusion models, multimodal learning, and recommender systems. He has authored numerous peer-reviewed publications and patented inventions spanning intelligent audio systems, voice technologies, and applied machine learning for consumer electronics. He was the recipient of the IEEE Ganesh N. Ramaswamy Memorial Student Grant at ICASSP 2015.
\end{IEEEbiography}

\begin{IEEEbiography}[{\includegraphics[width=1in,height=1.25in,clip,keepaspectratio]{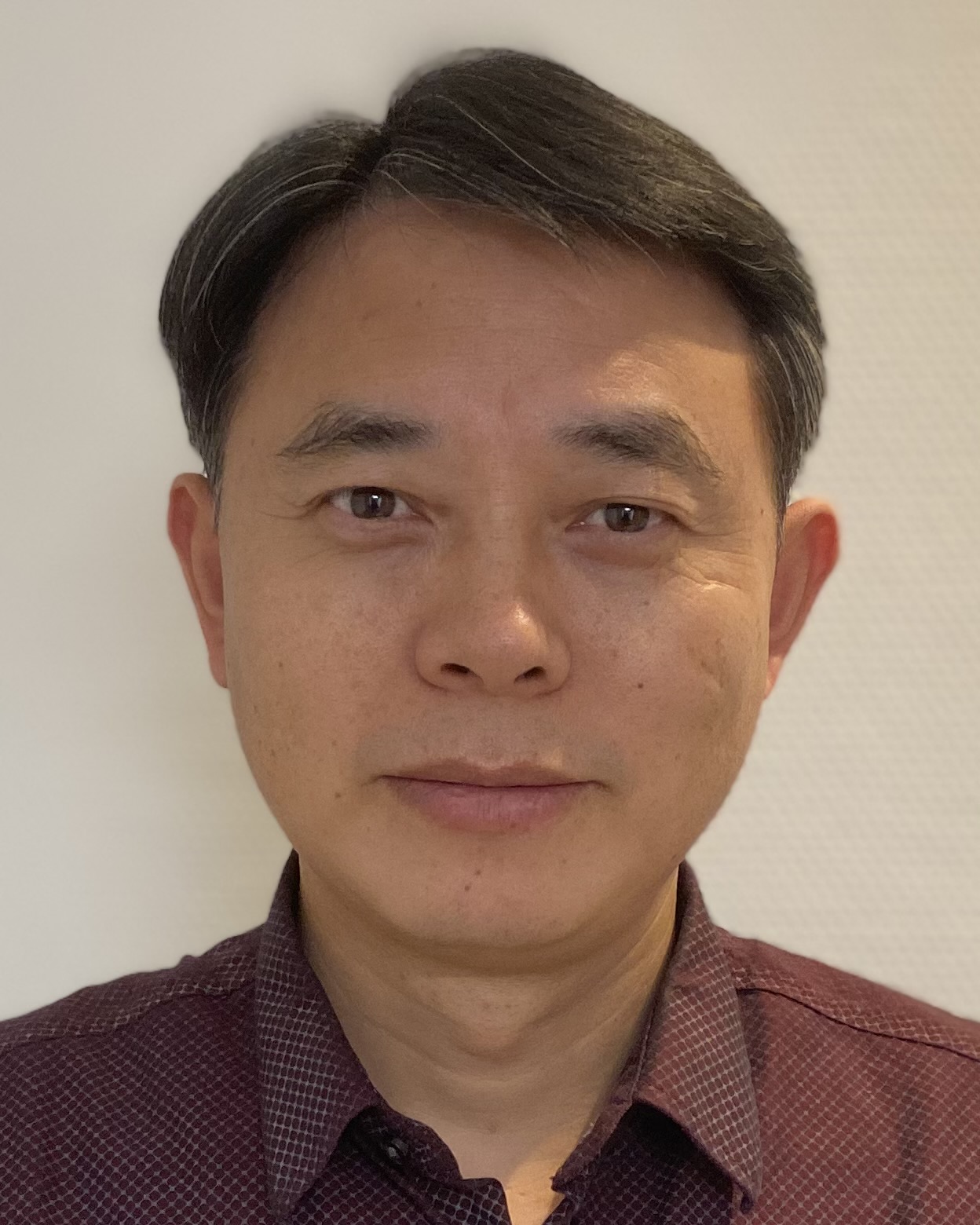}}]{Zheng-Hua Tan} received the B.Sc. and M.Sc. degrees in electrical engineering from Hunan University, Changsha, China, in 1990 and 1996, respectively, and the Ph.D. degree in electronic engineering from Shanghai Jiao Tong University (SJTU), Shanghai, China, in 1999. He is currently a Professor in the Department of Electronic Systems at Aalborg University, Aalborg, Denmark, and a Co-Head of the Centre for Acoustic Signal Processing Research. He is also a Co-Lead of the Pioneer Centre for AI, Denmark. He was a Visiting Scientist at the Computer Science and Artificial Intelligence Laboratory (CSAIL), MIT, USA, an Associate Professor at the Department of Electronic Engineering, SJTU, China, and a Postdoctoral Fellow at the AI Laboratory, KAIST, Korea. His research interests include machine learning, deep learning, noise-robust speech processing, and multimodal signal processing. He has (co-)authored over 300 peer-reviewed papers and received the IEEE Signal Processing Society Best Paper Award and the International Speech Communication Association Best Research Paper Award in 2022. He is the Lead Editor of the IEEE Journal of Selected Topics in Signal Processing Special Series on AI in Signal and Data Science. He has served as Chair of the IEEE SPS Machine Learning for Signal Processing Technical Committee and as a member of the IEEE SPS Technical Directions Board and Conferences Board, and as an Associate Editor for the IEEE TRANSACTIONS ON AUDIO, SPEECH AND LANGUAGE PROCESSING. He is the General Chair of ICASSP 2029 and a TPC Co-Chair for ICASSP 2028. He previously served as a TPC Vice-Chair for ICASSP 2024 and as General Chair of IEEE MLSP 2018.

\end{IEEEbiography}

\end{document}